\documentclass[paper]{JHEP3}
\usepackage{epsfig}
\usepackage{amssymb}
\newcommand\nl{n_{\ell}}
\def    \be             {\begin{equation}}
\def    \ee             {\end{equation}}
\def    \ba             {\begin{eqnarray}}
\def    \ea             {\end{eqnarray}}

\def    \frac           #1#2{{#1 \over #2}}

\def    \bra#1          {\mbox{$\langle #1 |$}}
\def    \ket#1          {\mbox{$| #1 \rangle$}}

\newcommand\MCatNLO{MC@NLO}
\newcommand\pt{p_{\scriptscriptstyle \rm T}}
\newcommand\kt{k_{\scriptscriptstyle \rm T}}

\def    \as             {\ifmmode\alpha_s\else$\alpha_s$\fi}
\def    \az             {\ifmmode\alpha_s^0\else$\alpha_s^0$\fi}

\def \sss{\scriptscriptstyle}
\def\mq{m_{\sss Q}}
\def\mQQ{M_{\sss Q\bar{Q}}}
\def\yqq{Y_{\sss Q\bar{Q}}}

\newcommand\Nc{N_c}

\newcommand\qb{\overline{q}}

\newcommand\fqq{f_{q\qb}}

\newcommand\fgg{f_{gg}}

\newcommand\fqg{f_{qg}}
\newcommand\fgqb{f_{g\bar{q}}}

\newcommand\thu{\theta_1}
\newcommand\thd{\theta_2}
\newcommand\omxplus{\left(\frac{1}{1-x}\right)_\rho}
\newcommand\omx{\left(\frac{1}{1-x}\right)}
\newcommand\omyplus{\left(\frac{1}{1-y}\right)_+}
\newcommand\opyplus{\left(\frac{1}{1+y}\right)_+}
\newcommand\ompyplus{\left(\frac{1}{1\mp y}\right)_+}
\newcommand\lomxplus{\left(\frac{\log(1-x)}{1-x}\right)_\rho}

\newcommand\opmyplus{\left(\frac{1}{1\pm y}\right)_+}
\def\abs#1{\left|#1\right|} 
\newcommand\cf{C_{\sss\rm F}}
\newcommand\tf{T_{\sss\rm F}}
\newcommand\ca{C_{\sss\rm A}}
\newcommand\da{D_{\sss\rm A}}

\newcommand\POWHEG{POWHEG}

\title{A Positive-Weight Next-to-Leading-Order Monte Carlo for
 Heavy Flavour Hadroproduction}
\vfill
\author{Stefano Frixione \\
INFN, Sezione di Genova, Italy\\
E-mail: \email{Stefano.Frixione@ge.infn.it}}
\author{Paolo Nason \\
INFN, Sezione di Milano Bicocca, Italy\\
E-mail: \email{Paolo.Nason@mib.infn.it}}
\author{Giovanni Ridolfi \\
Dipartimento di Fisica, Universit\`a di Genova\\
and INFN, Sezione di Genova, Italy\\
E-mail: \email{Giovanni.Ridolfi@ge.infn.it}}
\abstract{
We present a next-to-leading order calculation of heavy flavour production
in hadronic collisions that can be interfaced to shower Monte
Carlo programs. The calculation is performed in the context of the
POWHEG method~\cite{Nason:2004rx}.
It is suitable for the computation of charm, bottom and
top hadroproduction.  In the case of top production,
spin correlations in the decay products are taken into account.
}
\preprint{Bicocca-FT-07-12 \\GEF-TH-19/2007\\
}

\begin{document}
\section{Introduction}

The study of heavy flavoured hadronic final states produced in hadron
collisions has received great attention in the past twenty years. This
interest was driven partly by the search for the top quark, and partly
by the circumstance that processes characterized by one single large
energy scale (the heavy quark mass in this case) are reasonably
controlled by perturbative QCD due to asymptotic freedom. Theoretical
progress has therefore parallelled the experimental efforts in this
direction, and has by now reached a remarkable level of refinement.

The next-to-leading order (NLO) calculation of heavy flavour
hadroproduction has been matched to the HERWIG SMC
generator~\cite{Corcella:2000bw,Corcella:2002jc}
in the context of the MC@NLO 
formalism~\cite{Frixione:2002ik,Frixione:2003ei}.
This development is particularly important, since it has always
proved difficult to correctly simulate heavy flavour production
in standard shower Monte Carlo programs (SMC from now on),
because of the impact of higher
order processes like flavour excitation and gluon splitting.

In ref.~\cite{Nason:2004rx} a method for interfacing NLO calculations
with SMC generators was suggested, that overcomes some drawbacks of
the \MCatNLO\ technique. 
In particular, the implementation of a given production process with
the method of ref.~\cite{Nason:2004rx}
is independent of the SMC it will be matched to, whereas this
is not the case of \MCatNLO, which requires an SMC-dependent (but
process independent) contribution which is very laborious to obtain
(this is the reason why so far \MCatNLO\ has been matched only to
HERWIG).  On the other hand, the technique of ref.~\cite{Nason:2004rx}
requires the use of a soft shower
(the vetoed-truncated shower) which is not available in present SMC
generators, but whose effects are expected to be small. Furthermore,
the approach of ref.~\cite{Nason:2004rx} generates events with positive
weights, while \MCatNLO\ generates also events with negative weight.
For this reason the method has been referred to as \POWHEG{}, for POsitive
Weight Hard Event Generator.
In ref.~\cite{Nason:2006hf} the \POWHEG{} method was
successfully employed in the calculation of $Z$ pair production in
hadronic collisions.

In the present work, we apply the \POWHEG{} method to heavy flavour
hadroproduction. NLO cross section formulae for this process
have been available for a long time~\cite{Nason:1988xz,
Nason:1989zy,Mangano:1992jk}. In the application of the
\POWHEG{} method, no new analytic results are needed. Furthermore, since
the kinematics of the process is similar to that of $Z$ pair 
production~\cite{Nason:2006hf}, many results of that paper directly extend
to the case at hand. The result of our work provides a valuable alternative
to the \MCatNLO\ program. It can be used with both HERWIG and
PYTHIA~\cite{Sjostrand:2006za},
and it should not be difficult to interface it to other SMC's as well.
It furthermore provides a viable alternative, in cases when negatively
weighted events in \MCatNLO\ cause severe problems, like, for example,
when the heavy quark is relatively light compared to the energies involved.

The web location of the code repository can be found in the
manual, ref.~\cite{Frixione:2007nu}.

\section{Kinematics and cross section}\label{sec:kin}

The differential cross section for the production of heavy quark pairs
in hadronic collisions was computed in
ref.~\cite{Nason:1988xz},~\cite{Nason:1989zy}
and~\cite{Mangano:1992jk} up to order $\as^3$.  In this section, we
formulate the result of ref.~\cite{Mangano:1992jk} in a form which is
suitable for the generation of the hardest emission
using the procedure proposed in ref.~\cite{Nason:2004rx}.

The order-$\as^3$ cross section for the process $H_1 H_2\rightarrow
Q\bar{Q}+X$ can be written as the sum of four terms:
\be
d\sigma=d\sigma^{\rm (b)}+d\sigma^{\rm (sv)}+d\sigma^{\rm (f)}
+d\sigma^{\rm (c)}.
\ee
Here $d\sigma^{\rm (b)}$ is the leading-order (Born) cross section.  The
term $d\sigma^{\rm (sv)}$ collects order-$\as^3$ contributions with the same
two-body kinematics as the Born term, namely one-loop corrections
and real-emission contributions in the soft limit. Finally,
$d\sigma^{\rm (f)}$ represents the cross section for real emission, and
the corresponding subtractions, in a
generic configuration, while $d\sigma^{\rm (c)}$ is a remnant of the
subtraction of initial-state collinear singularities.

\subsection{Born and Soft-Virtual kinematics}
At leading order, the relevant parton subprocesses are
\ba
g(p_1)+g(p_2) &\rightarrow& Q(k_1)+ {\bar Q}(k_2),
\nonumber \\
q(p_1)+\bar q(p_2) &\rightarrow& Q(k_1)+ {\bar Q}(k_2),
\label{twobody}
\ea
where $q$ is a quark or antiquark of any flavour,
$\bar q$ the corresponding antiparticle, and $g$ is a gluon.
Particle four-momenta are displayed in brackets;
we have $p_1^2=p_2^2=0$, $k_1^2=k_2^2=\mq^2$, where $\mq$ is the
heavy quark mass. 

The kinematics of heavy quark pair production is entirely analogous
to that of vector boson pair production.
We introduce the reduced Mandelstam invariants\footnote{We depart  
slightly from the notation of ref.~\cite{Nason:2006hf}, 
where $t$ and $u$ have different definitions.}
\be
s=(p_1+p_2)^2,\qquad
t=(p_1-k_1)^2-\mq^2,\qquad
u=(p_1-k_2)^2-\mq^2,
\ee
related by $s+t+u=0$.

Event generation is conveniently performed in terms of the invariant mass
$\mQQ$ and the rapidity $\yqq$ of the $Q\bar Q$ pair
in the laboratory frame. They are given by
\ba
\label{M2xx}
&&\mQQ^2=(k_1+k_2)^2=x_1\,x_2\,S
\\
\label{Yxx}
&&\yqq=\frac{1}{2}\log\frac{(p_1+p_2)^0+(p_1+p_2)^3}
{(p_1+p_2)^0-(p_1+p_2)^3}=\frac{1}{2}\log\frac{x_1}{x_2},
\ea
where $S$ is the squared center-of-mass energy
of the colliding hadrons, and $x_1$, $x_2$ are the fractions of longitudinal
momenta carried by the incoming partons.
Equations~(\ref{M2xx},\ref{Yxx}) yield
\be
x_1=\sqrt{\frac{\mQQ^2}{S}}\,e^{\yqq}\equiv x_{b1};\quad
x_2=\sqrt{\frac{\mQQ^2}{S}}\,e^{-\yqq}\equiv x_{b2};\quad
dx_1 dx_2=\frac{1}{S}\,d\yqq d\mQQ^2.
\label{bornjac}
\ee
We adopt as two-body kinematic variables the set
$v=\{\mQQ,\yqq,\cos\theta_1\}$ (which we will call the Born variables
henceforth), where $\theta_1$ is the angle between
$\vec{p}_1$ and $\vec{k}_1$ in the partonic center-of-mass frame,
so that
\be
t=-\frac{\mQQ^2}{2}(1-\beta\cos\theta_1)\,,
\label{tborn}
\ee
where
\be
\beta=\sqrt{1-\rho};\qquad \rho=\frac{4\mq^2}{\mQQ^2}\,.
\label{betadef}
\ee
Using eq.~(\ref{bornjac}) and the usual expression
of the two-body phase space measure $d\Phi_2$, one
immediately finds
\be
d\Phi_2\,dx_1\,dx_2=
\frac{\beta}{16\pi S}\,d\cos\theta_1\,d\mQQ^2\,d\yqq\,.
\ee
In order to keep our notation similar to that of
ref.~\cite{Nason:2004rx}, we define
\be
\label{phiv}
d\Phi_v=d\cos\theta_1\,d\mQQ^2\,d\yqq.
\ee
The appropriate integration region for the variables $v$ is
\be
4\mq^2\leq \mQQ^2 \leq S\,,\quad
\frac{1}{2}\log\frac{\mQQ^2}{S}\leq \yqq\leq
-\frac{1}{2}\log\frac{\mQQ^2}{S}\,,\quad
-1\leq\cos\theta_1\leq 1.
\ee

The Born cross section is given by
\be
d\sigma^{\rm (b)}=d\Phi_v\,\sum_j B_j(v,\mu),
\ee
where
\be
\label{eq:bqdef}
B_j(v,\mu)=
\frac{\beta}{16\pi S}\,
f_j^{H_1}(x_{b1},\mu)\, f_{-j}^{H_2}(x_{b2},\mu)\,
{\cal M}_{j,-j}^{\rm (b)}(\mQQ^2,t).
\ee
The index $j$ represents the light quarks and antiquarks ($j\ne 0$)
and the gluon ($j=0$).  It ranges between $-\nl$ and $\nl$, where
$\nl$ is the number of light flavours.
In the following we will also use $q$ to represent all values of $j$ except for
$j=0$, $\bar q$ to represent $-j$, and $g$ to represent $j=0$.
$f_j^H(x,\mu)$ denotes
the distribution function of parton $j$ in the hadron $H$, and
$\mu$ is a factorization scale. The function 
${\cal M}_{j,-j}^{\rm (b)}(s,t)$ is the squared invariant amplitude,
summed over final-state polarizations and averaged over initial-state
polarizations and colours, divided by the relevant flux factor.
We have
\ba
&&{\cal M}_{gg}^{\rm (b)}(s,t)=
\frac{g^4}{2s}\frac{2 \tf}{\da}\left(\cf\frac{s}{ut}-\ca\right)
\left(\frac{t^2}{s^2}+\frac{u^2}{s^2}+\rho-\frac{\mq^4}{tu}\right)\,,
\nonumber \\
&& {\cal M}_{q\bar q}^{\rm (b)}(s,t)=
\frac{g^4}{2s}\frac{\cf\tf}{\Nc}
 \left(\frac{2t^2}{s^2}+\frac{2u^2}{s^2}+\rho\right)\,,
\label{Born}
\ea
where $\Nc=3$, $\tf=1/2$, $\cf=4/3$, $\da=8$.

Order-$\as^3$ contributions to the cross section arise from one-loop
corrections to the two-body process eq.~(\ref{twobody}), and from
real-emission subprocesses at tree level. The
contribution of one-loop diagrams must be summed to the one-gluon
emission cross section in the soft limit, in order to obtain an
infrared-finite result. The resulting contribution,
usually called the soft-virtual contribution,
has the same kinematic structure as the leading-order term:
\be
d\sigma^{\rm (sv)}=d\Phi_v\sum_j V_j(v,\mu),
\ee
where
\be
V_j(v,\mu)=\frac{\beta}{16\pi S}\,
f_j^{H_1}(x_{b1},\mu)\, f_{-j}^{H_2}(x_{b2},\mu)
{\cal M}^{\rm (sv)}_{j,-j}(\mQQ^2,t,\mu^2)\,.
\ee
The invariant amplitude ${\cal M}^{\rm (sv)}_{j,-j}(\mQQ^2,t,\mu^2)$
is the sum of the virtual corrections to the 2 body
subprocesses, and the soft contribution of real emission.
The sum is finite, and it was computed in ref.~\cite{Mangano:1992jk}.

\subsection{Real emission kinematics}
We consider now the real-emission subprocesses
\ba
q(p_1)+\bar q(p_2)&\rightarrow& Q(k_1)+\bar Q(k_2)+g(k)
\label{qq}
\\
q(p_1)+g(p_2)&\rightarrow& Q(k_1)+\bar Q(k_2)+q(k)
\label{qg}
\\
g(p_1)+\bar q(p_2)&\rightarrow& Q(k_1)+\bar Q(k_2)+\bar q(k)
\label{gq}
\ea
in a generic kinematical configuration.
The processes (\ref{qq}-\ref{gq})
are characterized by five independent scalar quantities, which
we choose to be
\ba
&&s=(p_1+p_2)^2,\quad
t_k=(p_1-k)^2, \quad
u_k=(p_2-k)^2, \quad
\\
&&q_1=(p_1-k_1)^2-\mq^2,\quad
q_2=(p_2-k_2)^2-\mq^2,
\ea
as in ref.~\cite{Mangano:1992jk}. We introduce the variables 
\be
x=\frac{\mQQ^2}{s};\qquad y=\cos\theta,
\ee
where $\theta$ is the scattering angle of the emitted parton
in the partonic center-of-mass system. With these definitions,
\be
t_k=-\frac{s}{2}(1-x)(1-y);\qquad
u_k=-\frac{s}{2}(1-x)(1+y).
\label{tandu}
\ee
It is easy to show that in the case of the subprocesses
(\ref{qq}-\ref{gq}) one has
\be
\yqq=\frac{1}{2}\log\left[\frac{x_1}{x_2}\frac{s+u_k}{s+t_k}\right]
=\frac{1}{2}\log\left[\frac{x_1}{x_2}\frac{2-(1-x)(1+y)}{2-(1-x)(1-y)}\right];
\qquad
\mQQ^2=x\,x_1x_2\,S,
\ee
and therefore
\be
x_1=\frac{x_{b1}}{\sqrt{x}}\,\sqrt{\frac{2-(1-x)(1-y)}{2-(1-x)(1+y)}};
\qquad
x_2=\frac{x_{b2}}{\sqrt{x}}\,\sqrt{\frac{2-(1-x)(1+y)}{2-(1-x)(1-y)}}
\label{x1x2}
\ee
and
\be
\label{mcjac}
dx_1\,dx_2=\frac{1}{xS}\,d\mQQ^2\,d\yqq.
\ee
The range for the variable $x$ is restricted by the requirement
that both $x_1$ and $x_2$ be smaller than one; this gives
\be
x_{\rm min}\leq x\leq 1,
\ee
with
\ba
x_{\rm min}&=& {\rm max}\left(
\frac{2(1+y)\,x_{b1}^2}{\sqrt{(1+x_{b1}^2)^2(1-y)^2+16yx_{b1}^2}
    +(1-y)(1-x_{b1}^2)},\right.
\nonumber\\
&&\phantom{aaaaa}\left.
\frac{2(1-y)\,x_{b2}^2}{\sqrt{(1+x_{b2}^2)^2(1+y)^2-16yx_{b2}^2}
    +(1+y)(1-x_{b2}^2)}\right).
\label{xmin}
\ea
Note that $x_{\rm min}$ depends explicitly on $y$,
and implicitly on $\mQQ^2$ and $\yqq$ through $x_{b1},x_{b2}$.
It can be checked that $x_{\rm min}$ is always larger that $\mQQ^2/S$,
as required by the definition of~$x$.

In the center-of-mass frame of the $Q\bar{Q}$ system,
the four-momenta of the produced heavy quarks can be parametrized
in terms of two angles $\theta_1,\theta_2$:
\ba
k_1&=&\frac{\mQQ}{2}\;
(1,\beta\sin\theta_2\sin\theta_1,
\beta\cos\theta_2\sin\theta_1,\beta\cos\theta_1) \nonumber \\
k_2&=&\frac{\mQQ}{2}\;
(1,-\beta\sin\theta_2\sin\theta_1,
-\beta\cos\theta_2\sin\theta_1,-\beta\cos\theta_1),
\ea
with $\beta$ given in eq.~(\ref{betadef}).
Both $\theta_1$ and $\theta_2$ range between $0$ and $\pi$.
Thus, in addition to the Born variables $v$, 
we have now the three radiation variables $r=\{x,y,\thd\}$, with
\be
x_{\rm min}\leq x\leq 1\,,\quad
-1\leq y\leq 1\,,\quad
0\leq\thd\leq\pi\,.
\ee
Following ref.~\cite{Nason:2004rx}
we define the corresponding integration measure
\be
\label{dphir}
d\Phi_r=dx\,dy\,d\thd.
\ee
From the computation of ref.~\cite{Mangano:1992jk} we obtain
\be
d\sigma^{\rm (f)}=d\Phi_v\,d\Phi_r\,
\left\{ R_{gg}(v,r,\mu)+\sum_q \left[R_{q\bar q}(v,r,\mu)+R_{qg}(v,r,\mu)
+R_{g\bar q}(v,r,\mu)\right]\right\},
\ee
where
\ba
R_{gg}(v,r,\mu)&=&
\frac{1}{(4\pi)^2}
\frac{\beta}{64\pi^2 \mQQ^2 S}\,
\omxplus\left[\omyplus+\opyplus\right]
\label{eq:rdef}
\\
&&f_g^{H_1}(x_1,\mu)\, f_{g}^{H_2}(x_2,\mu)\,
\fgg(x,y,\thu,\thd,\mu)
\nonumber\\
R_{q\bar q}(v,r,\mu)&=&
\frac{1}{(4\pi)^2}
\frac{\beta}{64\pi^2 \mQQ^2 S}\,
\omxplus\left[\omyplus+\opyplus\right]
\nonumber\\
&&f_q^{H_1}(x_1,\mu)\, f_{\bar q}^{H_2}(x_2,\mu)\,
\fqq(x,y,\thu,\thd,\mu)
\nonumber\\
R_{qg}^\pm(v,r,\mu)&=&
\frac{1}{(4\pi)^2}
\frac{\beta}{64\pi^2 \mQQ^2 S} \omx
\ompyplus f_q^{H_1}(x_1,\mu)\, f_g^{H_2}(x_2,\mu)\,
\fqg(x,y,\thu,\thd,\mu)
\nonumber\\
R^\pm_{g\bar q}(v,r,\mu)&=&
\frac{1}{(4\pi)^2}
\frac{\beta}{64\pi^2 \mQQ^2 S} \omx
\ompyplus f_g^{H_1}(x_1,\mu)\, f_{\bar q}^{H_2}(x_2,\mu)\,
\fgqb(x,y,\thu,\thd,\mu).
\nonumber
\ea
The functions 
$R_{q\bar q}$, $R_{qg}$, $R_{g\bar q}$
denote the regularized
real emission cross sections for the different subprocesses.
The functions $\fqq$ and $\fqg$ are regular in the limits
of soft ($x=1$) or collinear ($y=\pm 1$) emission;
they are defined as in eq.~(3.3) of ref.~\cite{Mangano:1992jk}.
The distributions $1/(1-x)_\rho$ and $1/(1\pm y)_+$
are defined by
\ba
\label{omxdef}
&&\int_\rho^1dx\,g(x)\,\omxplus=
\int_\rho^1dx\,\frac{g(x)-g(1)}{1-x}
\\
&&\int_{-1}^1dy\,h(y)\,\opmyplus=
\int_{-1}^1dy\,\frac{h(y)-h(\mp 1)}{1\pm y}.
\label{opmydef}
\ea

\subsection{Collinear remnants}
The remnants of the collinear subtraction must also be added to get the
full cross section. This contribution has the form~\cite{Mangano:1992jk}
\ba
d\sigma^{\rm (c)}&=&d\Phi_v\,dx\,dy\,
 \\&\Big\{ &
\Big[ L^+_{gg}(v,x,\mu)+\sum_q
\left( L^+_{q\bar q}(v,x,\mu)+L^+_{g\bar q}(v,x,\mu) +L^+_{qg}(v,x,\mu) \right) \Big]\,\delta(1-y)
\nonumber \\\nonumber
& +&\Big[L^-_{gg}(v,x,\mu)+\sum_q \left(L^-_{q\bar q}(v,x,\mu)+L^-_{g\bar q}(v,x,\mu)
+L^-_{qg}(v,x,\mu)\right) \Big]\,\delta(1+y)\;\;\Big\}\,,
\ea
where
\ba
L^+_{ij}(x) &=&\frac{\as}{2\pi} \frac{\beta}{16\pi S}
 \sum_{i'} \Bigg\{\left[ (1-x) P_{i i'}(x,0) \right]
\Bigg[\omxplus\log\frac{\mQQ^2}{x\mu^2}
+2\lomxplus\Bigg] \phantom{AAA}
\nonumber \\*&&
-\left[ (1-x) P^\prime_{i i'}(x,0) \right] \omxplus
\Bigg\} \, {\cal M}_{i' j}^{\rm (b)}(\mQQ^2,t)
f_i^{H_1}(x_1,\mu)\,f_j^{H_2}(x_2,\mu)
\label{eq:FKScrp}
\ea
and
\ba
L^-_{ij}(x) &=&\frac{\as}{2\pi} \frac{\beta}{16\pi S}
 \sum_{j'} \Bigg\{\left[ (1-x) P_{j j'}(x,0) \right]
\Bigg[\omxplus\log\frac{\mQQ^2}{x\mu^2}
+2\lomxplus\Bigg] \phantom{AAA}
\nonumber \\*&&
-\left[ (1-x) P^\prime_{j j'}(x,0) \right] \omxplus
\Bigg\} \, {\cal M}_{i j'}^{\rm (b)}(\mQQ^2,t)
f_i^{H_1}(x_1,\mu)\,f_j^{H_2}(x_2,\mu)
\label{eq:FKScrm}
\ea
where $P_{ij}(x,\epsilon)$ are the leading-order Altarelli-Parisi
splitting functions in $d=4-2\epsilon$ dimensions,
$P^\prime_{ij}(x,\epsilon)$ their first derivatives with respect to
$\epsilon$,
and $t$ is given in eq.~(\ref{tborn}).
From eq.~(\ref{xmin}) we see that
the integration range becomes
$x_{b1}<x<1$ when $y=1$,
and $x_{b2}<x<1$ for $y=-1$.

\section{Cross section for the hardest emission}\label{sec:hard}
\begin{figure}[ht]
\begin{center}
\epsfig{file=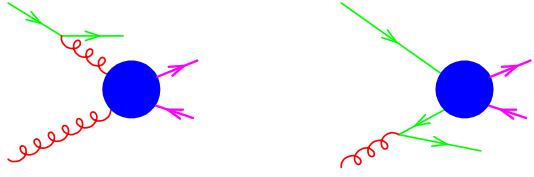,width=7cm}
\end{center}
\caption{\label{fig:lump}
Different underlying Born structure in the two collinear regions of the
$qg$ subprocess.}
\end{figure}

The POWHEG method, when applied to a generic
process, may require a separated treatment of each singular region.
In the present case (as in the case of $Z$ pair production) this is not needed.
Our choice of variables $v,r$ is adequate for both collinear
regions at the same time, the only difference being the sign of $y$.
We have instead to pay attention to the
flavour structure of the process.
In ordinary SMC codes, the flavour structure of the Born
subprocess is not altered by subsequent radiation.
On the other hand, if
the hardest radiation is produced in the context of a NLO
calculation, the association of the NLO process with
a Born subprocess is not always obvious. A given real-emission
contribution must be associated with its underlying Born process, i.e.
the Born process in which it factorizes in the collinear limit.
In the present case, the collinear regions for the real $q\bar{q}$
subprocess always factorize in terms of the $q\bar{q}$ underlying Born,
and the collinear regions for the real $gg$ subprocess factorize in terms 
of the $gg$ Born process. For the $qg$ ($g\bar{q}$) processes, there are
instead two possibilities (see fig.~\ref{fig:lump}): the underlying
Born process is $gg$ ($q\bar{q}$) for the $y=1$ collinear region, and 
$q\bar{q}$ ($gg$) for $y=-1$. 
This is the reason why in eq. (\ref{eq:rdef}) we have separated
the two collinear contributions $R_{qg}^\pm$ and $R^\pm_{g\bar q}$.
Thus, for a given flavour~$q$, we lump together the
$q\bar{q}$, the $qg$ and the $g\bar{q}$ real-emission subprocesses.

Following ref.~\cite{Nason:2004rx}, we write the cross section for
the event with the hardest emission as
\begin{eqnarray}
&&d\sigma = \sum_q {\bar B}_q(v,\mu_v) d\Phi_v \left[
\Delta_q(0)+\Delta_q(\kt)
\frac{{\hat R}_{q\bar{q}}(v,r,\mu_r)+{\hat R}^-_{qg}(v,r,\mu_r)
+{\hat R}^+_{g\bar{q}}(v,r,\mu_r)}{B_q(v,\mu_r)}
d\Phi_r \right]\, \nonumber 
\\*&&
+ {\bar B}_g(v,\mu_v) d\Phi_v \left[
\Delta_g(0)+\Delta_g(\kt)
\frac{{\hat R}_{gg}(v,r,\mu_r)+\sum_q\left({\hat R}^+_{qg}(v,r,\mu_r)
+{\hat R}^-_{g\bar{q}}(v,r,\mu_r)\right)}{B_q(v,\mu_r)}
d\Phi_r \right]\,,
\nonumber\\*
\label{eq:hardestsigma}
\end{eqnarray}
where $\hat{R}_{ij}$ is obtained from $R_{ij}$ by removing
the $+$ prescriptions that regularize the $x$ and $y$ singularities.
The $\hat{R}_{ij}$ are thus the unregularized real emission cross sections
(corresponding to $R$ in the notation of
ref.~\cite{Nason:2004rx}). Notice that the real emission contributions
having the same underlying Born configuration are grouped together in
eq.~(\ref{eq:hardestsigma}).
Furthermore,
\ba
{\bar B}_q(v,\mu) &=& 
B_q(v,\mu)+V_q(v,\mu)+\int d\Phi_r \,
\left[ R_{q\bar{q}}(v,r,\mu)+R^-_{qg}(v,r,\mu)
+R^+_{g\bar{q}}(v,r,\mu)\right]
\nonumber \\
&&+\int_{-1}^1 dy \int_{x_{\rm min}}^1 dx\,
\left[L^+_{q\bar{q}}(v,x,\mu)+L^+_{g\bar{q}}(v,x,\mu)\right]\,\delta(1-y)
\nonumber \\ \label{eq:bbqdef}
&&+\int_{-1}^1 dy \int_{x_{\rm min}}^1 dx\,
\left[L^-_{q\bar{q}}(v,x,\mu)+L^-_{qg}(v,x,\mu)\right]\,\delta(1+y)
\ea
\ba
{\bar B}_g(v,\mu) &=& 
B_g(v,\mu)+V_g(v,\mu)+\int d\Phi_r \,
\left[ R_{gg}(v,r,\mu)+\sum_q \left( R^+_{qg}(v,r,\mu)
+R^-_{g\bar{q}}(v,r,\mu)\right)\right]
\nonumber \\
&&+\int_{-1}^1 dy \int_{x_{\rm min}}^1 dx\,
\left[L^+_{gg}(v,x,\mu)+\sum_q L^+_{qg}(v,x,\mu)\right]\,\delta(1-y)
\nonumber \\ \label{eq:bbgdef}
&&+\int_{-1}^1 dy \int_{x_{\rm min}}^1 dx\,
\left[L^-_{q\bar{q}}(v,x,\mu)
+\sum_q L^-_{g\bar{q}}(v,x,\mu)\right]\,\delta(1+y)
\ea

\be
\label{eq:Deltaq}
\Delta_q(\pt)=\exp \left[ - \int
\frac{{\hat R}_{q\bar{q}}(v,r,\mu_r)+{\hat R}^-_{qg}(v,r,\mu_r)
+{\hat R}^+_{g\bar{q}}(v,r,\mu_r)}{B_q(v,\mu_r)} \theta(\kt(v,r)-\pt)
 d\Phi_r\right]\,,
\ee
\be
\label{eq:Deltag}
\Delta_g(\pt)=\exp \left[ - \int
\frac{{\hat R}_{gg}(v,r,\mu_r)+\sum_q {\hat R}^+_{qg}(v,r,\mu_r)
+\sum_q {\hat R}^-_{g\bar{q}}(v,r,\mu_r)}{B_g(v,\mu_r)} \theta(\kt(v,r)-\pt)
 d\Phi_r\right]\,,
\ee
and $\kt(v,r)$ is the transverse momentum of the radiated parton,
\begin{equation} \label{eq:ktdef}
\kt(v,r)=\sqrt{\frac{\mQQ^2}{4x}(1-x)^2\,(1-y^2)}\;.
\end{equation}
Equation~(\ref{eq:hardestsigma}) is the analogue of eq.~(5.10)
of ref.~\cite{Nason:2004rx}. The function $\Delta_q(\pt)$ corresponds to
$\Delta_R^{(\rm NLO)}(\pt)$ in the notation of ref.~\cite{Nason:2004rx}.

\section{Generation of the hardest event}
The generation of the hardest event according to eq.~(\ref{eq:hardestsigma})
can be performed in full analogy to the case of ref.~\cite{Nason:2006hf}.
We refer the reader to that paper for details. Here we only point out
the relevant differences with respect to that case.

The generation of the Born configuration involves in this case two kinds
of Born processes, the $q\bar{q}$ and $gg$ processes. The total cross
section is thus given by
\begin{equation}
\sigma_{\rm tot}=\int d\Phi_v\,\left[{\bar B}_g(v,\mu_v) + 
\sum_q {\bar B}_q(v,\mu_v)\right]\,.
\end{equation}
After the Born configuration has been generated, one chooses the 
process (i.e. $g$
or a given flavour of $q$) with a probability proportional to 
${\bar B}_g(v,\mu_v)$,
${\bar B}_q(v,\mu_v)$. According to whether a $q$ or a $g$ was selected,
one follows the same procedure of ref.~\cite{Nason:2006hf} using to
the first or second line of eq.~(\ref{eq:hardestsigma}) respectively.

\section{Accuracy of the Sudakov form factor}
Unlike the case of $Z$ pair production, in the case at hand the
procedure illustrated in section~4 of ref.~\cite{Nason:2006hf} 
(i.e. the redefinition of $\as$ given in eq. (4.9) of~\cite{Nason:2006hf})
is not
sufficient to guarantee full next-to-leading logarithmic
accuracy of the Sudakov form
factor. This is related to the fact that the heavy flavour production
process at the Born level involves more than 3 coloured
partons~\cite{Bonciani:2003nt}, so that soft emission cross sections
do not simply factorize in terms of the Born cross section. 
Thus, the Sudakov form factor is strictly only accurate to leading log.
In fact, next-to-leading logarithmic
accuracy can be easily recovered at least in the large $N_c$
limit, where $N_c$ is the number of colours, with a procedure
discussed in ref.~\cite{Oleari:2007zz}. The implementation of this
procedure and the assessment of its impact is left for future work.

\section{Colour assignment}

In order to interface \POWHEG{} with a shower Monte Carlo,
colour connections must be specified. In the case of $Z$ pair
production, only one colour structure is present.
The situation is more complex in the case at hand, since
more colour structures are relevant. This problem is dealt with in
exactly the same way as was done in ref.~\cite{Frixione:2003ei},
section 6.1. We used in fact the same decomposition of the large $N_c$
heavy flavour production cross section into contributions with
different colour structure, and pick the colour structure with
probability proportional to the value of the corresponding
contribution.

\section{Results}

In this section, we present results obtained with the POWHEG
method for a choice of observables relevant to heavy quark
production. Our results will be compared to those obtained with
\MCatNLO. We thus interface POWHEG with HERWIG, in order to
make a consistent comparison.
The formalisms of POWHEG and \MCatNLO\ differ in the treatment
of contributions 
of orders higher than NLO, which are beyond the level of accuracy 
of the theoretical computations presently available.
The difference is mainly due to the way the radiation of matrix element origin
is generated, which is typically 
the hardest radiation.
Furthermore, scale choices in the two codes are not the same.
Therefore, sizable differences between the two methods are to be expected 
for bottom and charm production, where the relevant scale is relatively
small, while in the case of top production the discrepancies should in
principle be much less important.  However, a detailed comparison
between the two methods is beyond our present purposes, and it is left
for future work.

We will consider two experimental configurations: $p\bar p$
collisions at \mbox{$\sqrt{s}=1960$~GeV}, corresponding to the Tevatron 
Run II configuration, and $pp$ collisions at \mbox{$\sqrt{s}=14$~TeV}, 
corresponding to the LHC. The results presented in this section have
been obtained by setting the top and bottom masses equal to $172$~GeV
and $4.75$~GeV respectively. We have used the MRST2002~\cite{Martin:2002aw}
set of parton distributions. When considering
the decay of top quarks, we have set $\Gamma_t=1.31$~GeV.

\begin{figure}[ht]
\begin{center}
\epsfig{file=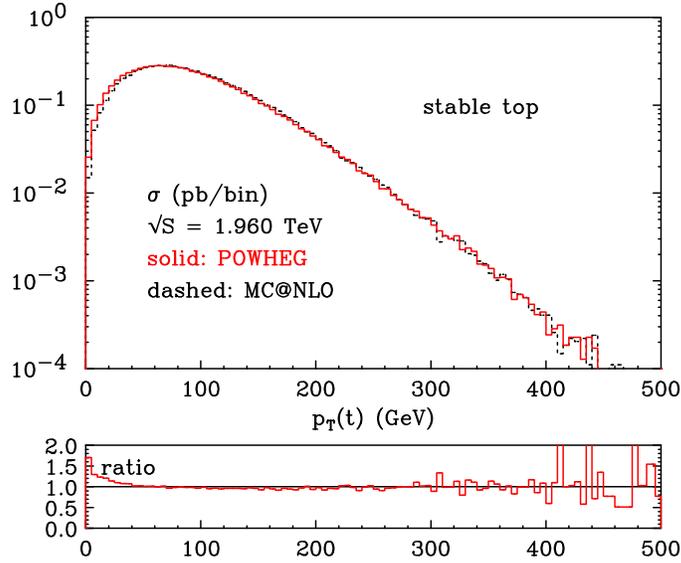,width=9cm}
\end{center}
\caption{\label{fig:toptevsta1}
Transverse momentum distribution of a top quark at the Tevatron.}
\end{figure}

We begin by considering top production. Both the POWHEG
and \MCatNLO\ codes include the possibility of generating distributions
for either undecayed top quarks (which we will refer to as ``stable top''
in the following), or for the decay products of 
top quarks, taking spin correlations into account.
We will show examples of both cases.
We present in fig.~\ref{fig:toptevsta1} the single-inclusive
transverse momentum distribution for a stable top quark produced
at the Tevatron.
The ratio between the POWHEG and the \MCatNLO\ results is also shown
in the lower pane.
The POWHEG (solid histogram) and \MCatNLO\ (dashed histogram) results
are very close to each other over the whole range considered,
except in the very small $\pt$ region, where the POWHEG cross section 
tends to be larger.

\begin{figure}[ht]
\begin{center}
\epsfig{file=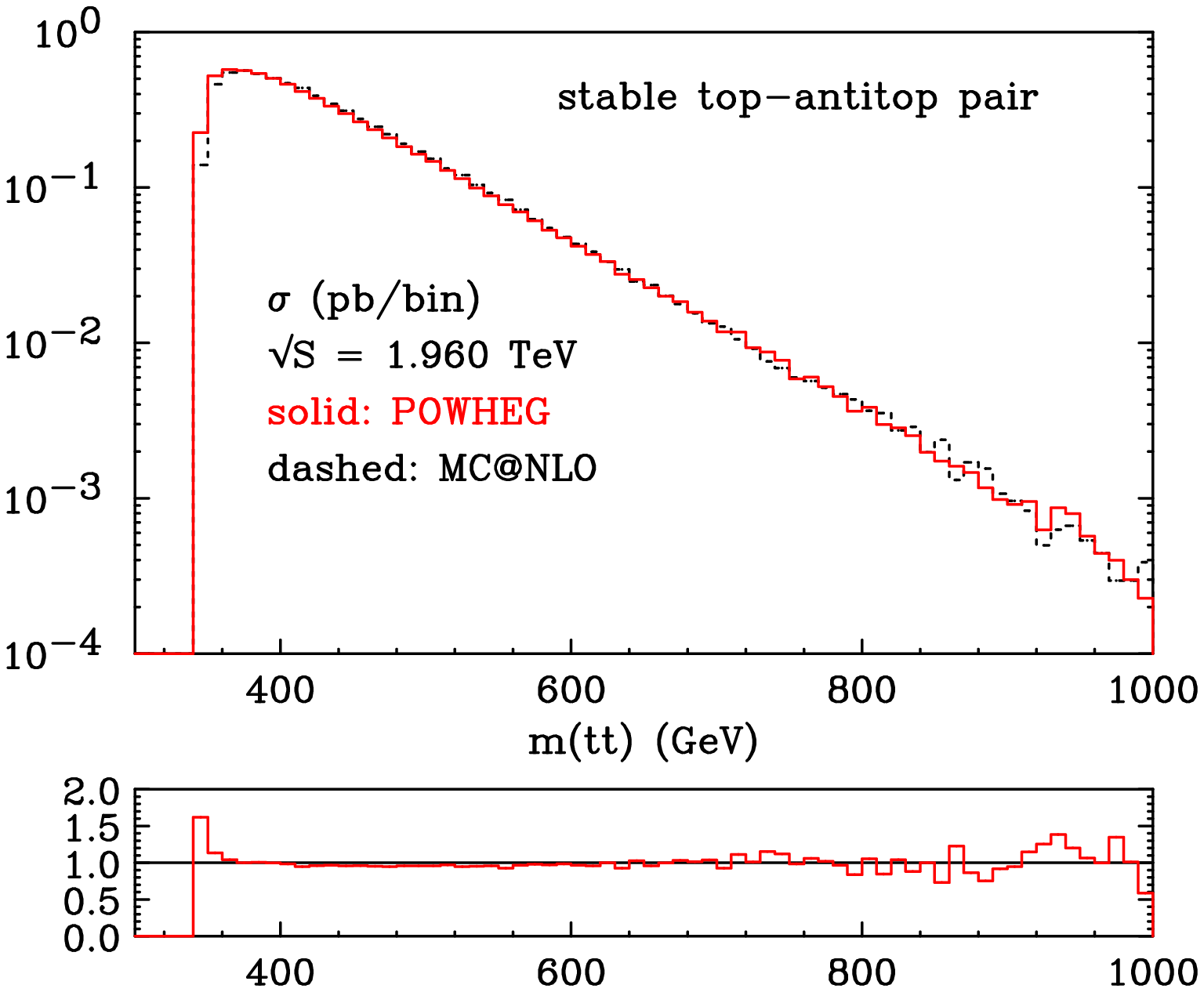,width=7cm}
\epsfig{file=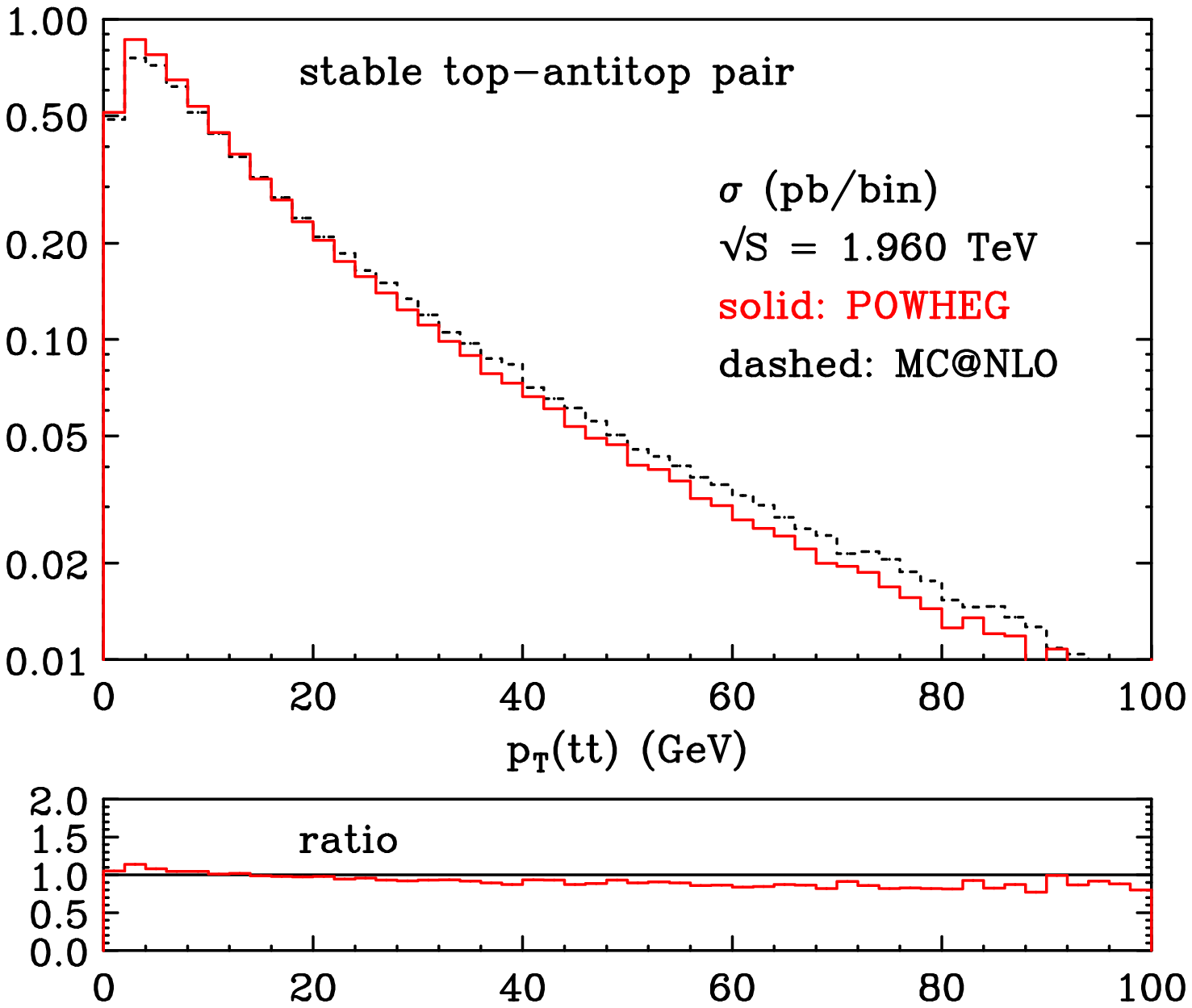,width=7cm}
\end{center}
\caption{\label{fig:toptevsta2}
Invariant mass and transverse momentum distributions of $t\bar t$ pairs
at the Tevatron.}
\end{figure}

In fig.~\ref{fig:toptevsta2} we present the distributions of the 
invariant mass and of the transverse momentum of a stable $t\bar t$ pair.
The agreement in this case is also quite good;
the POWHEG $\pt(t\bar t)$ distribution is slightly softer than that of
\MCatNLO.

\begin{figure}[ht]
\begin{center}
\epsfig{file=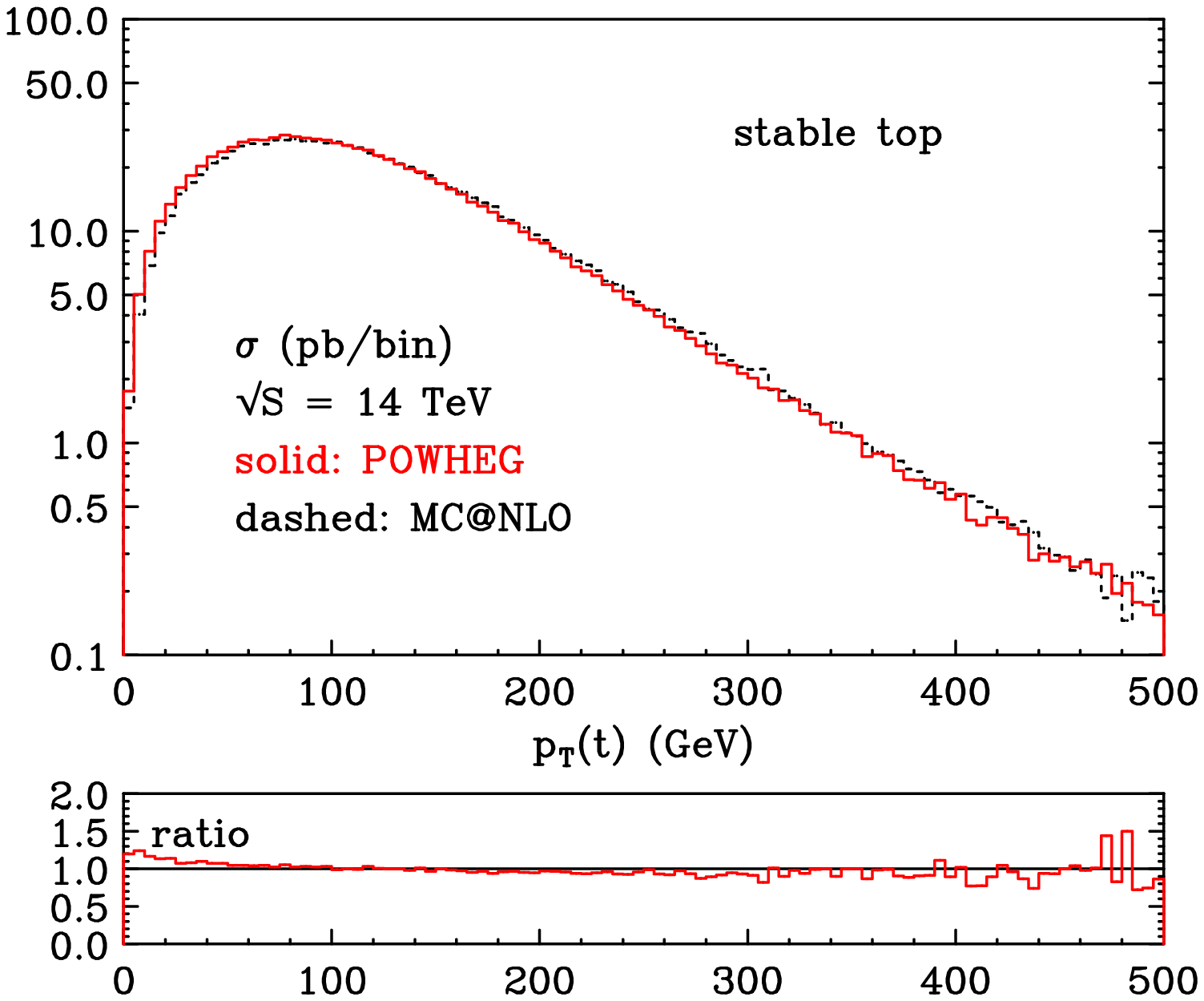,width=9cm}
\end{center}
\caption{\label{fig:toplhcsta1}
Transverse momentum distribution of a top quark at the LHC.}
\end{figure}

\begin{figure}[ht]
\begin{center}
\epsfig{file=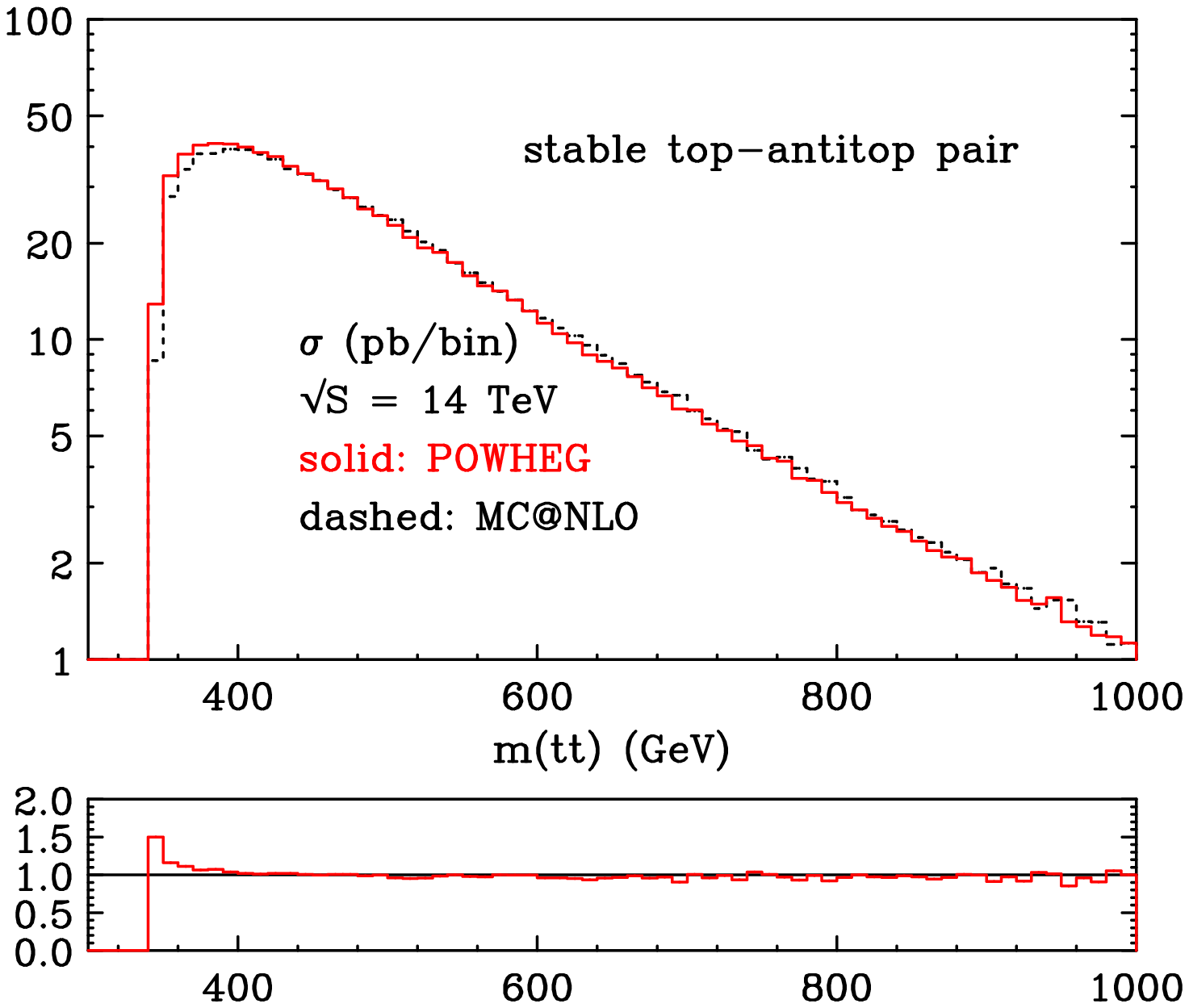,width=7cm}
\epsfig{file=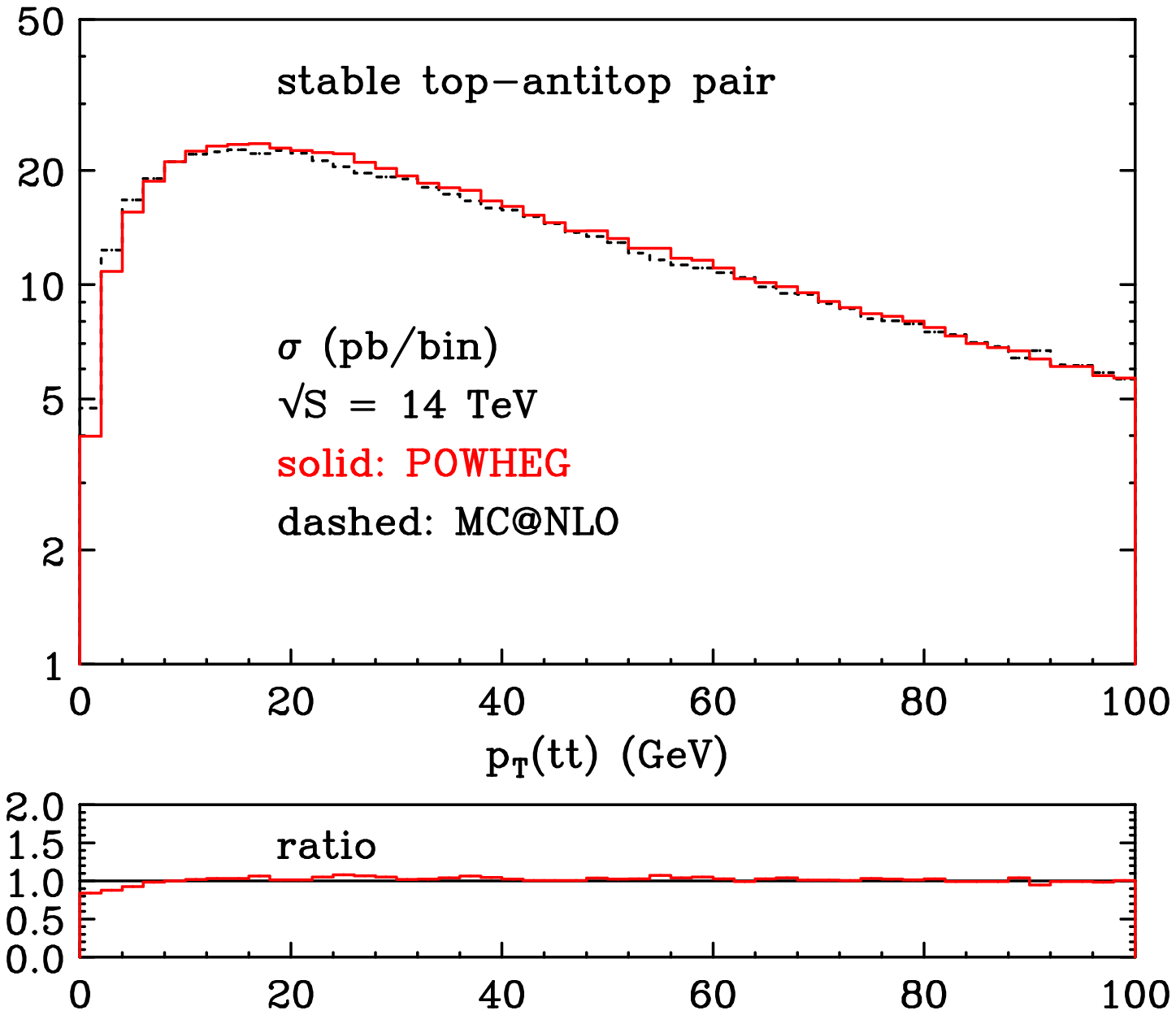,width=7cm}
\end{center}
\caption{\label{fig:toplhcsta2}
Invariant mass and transverse momentum distributions of $t\bar t$ pairs
at the LHC.}
\end{figure}

The same observables are in an even better agreement in the case 
of the LHC. This is shown in figs.~\ref{fig:toplhcsta1}
and~\ref{fig:toplhcsta2}.
This can be understood as a consequence of the fact that the
kinematics of the production process is less constrained than
at the Tevatron, therefore making the contribution of potentially
large logarithms in the perturbative coefficients less important.

\begin{figure}[ht]
\begin{center}
\epsfig{file=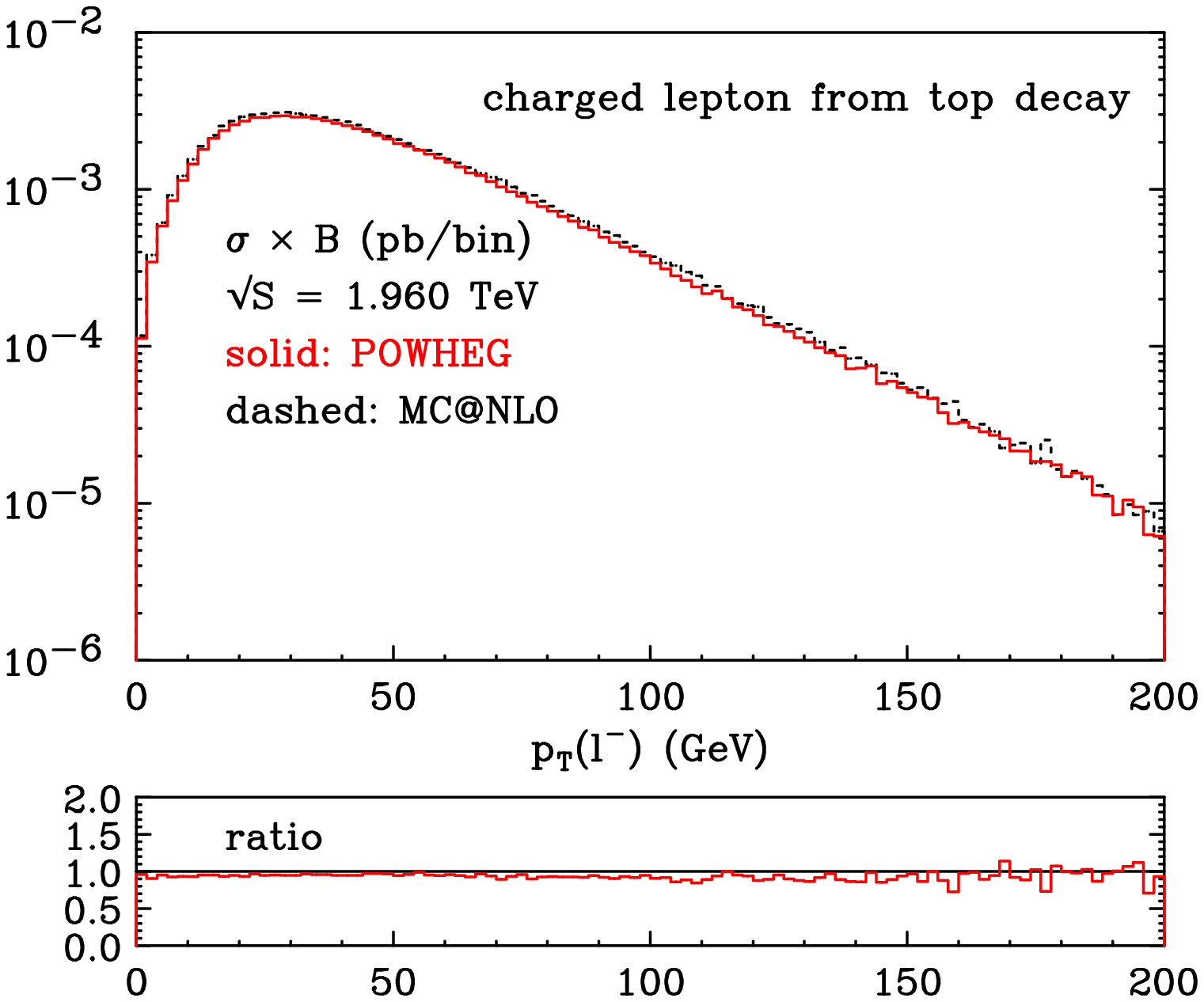,width=7cm}
\epsfig{file=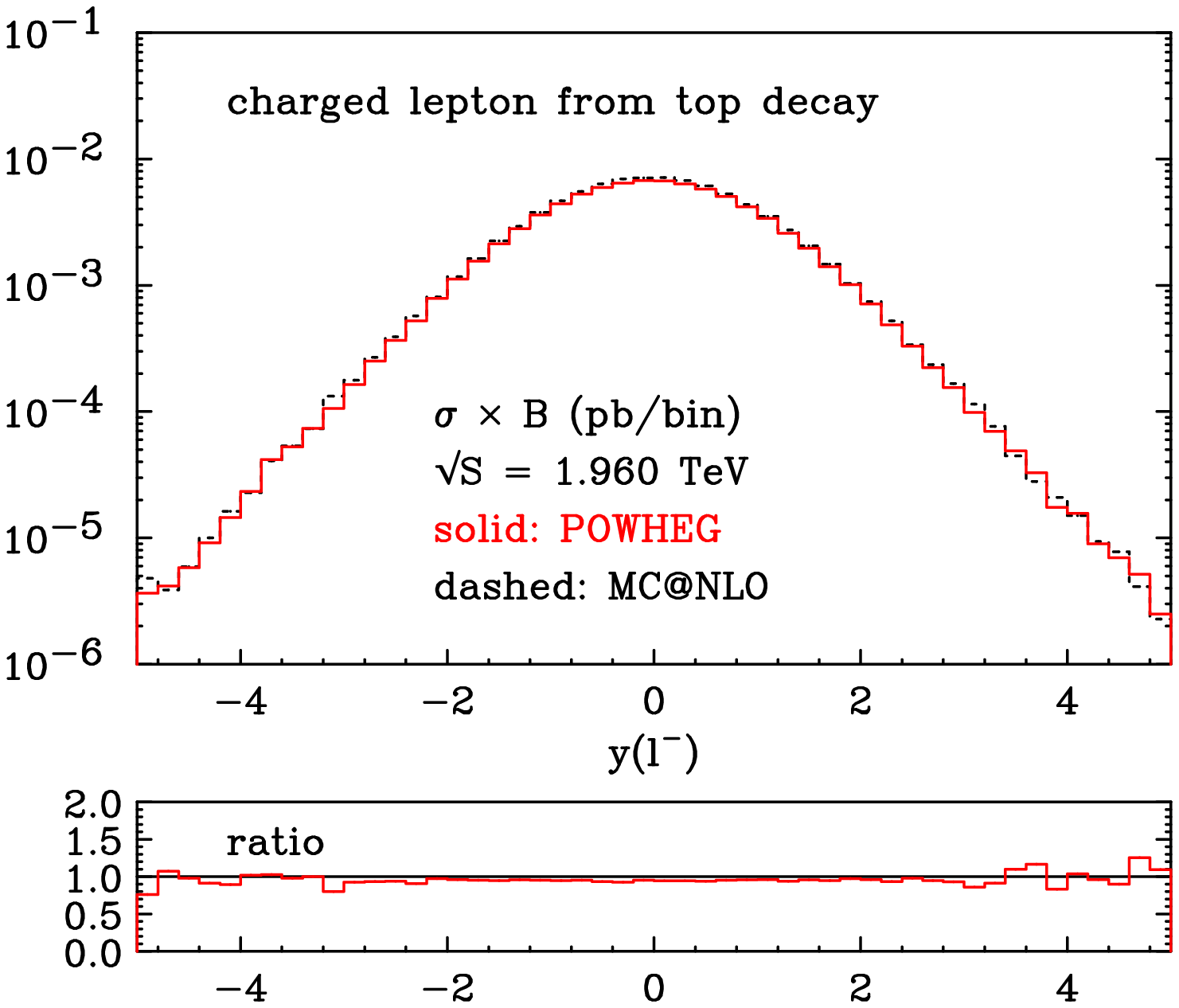,width=7cm}
\end{center}
\caption{\label{fig:toptevdec1}
Transverse momentum and rapidity of a charged lepton from top decay
at the Tevatron.}
\end{figure}

\begin{figure}[ht]
\begin{center}
\epsfig{file=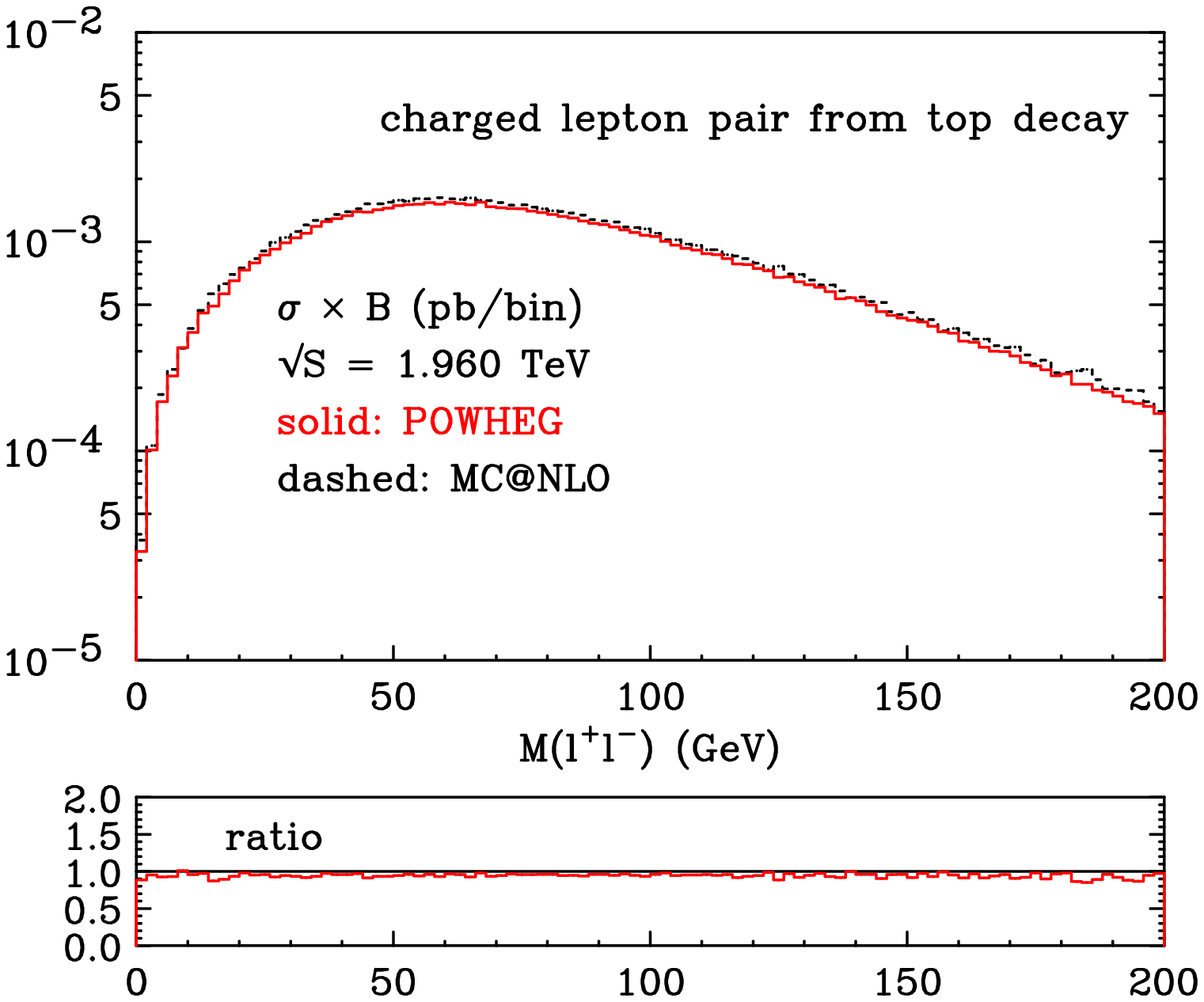,width=7cm}
\epsfig{file=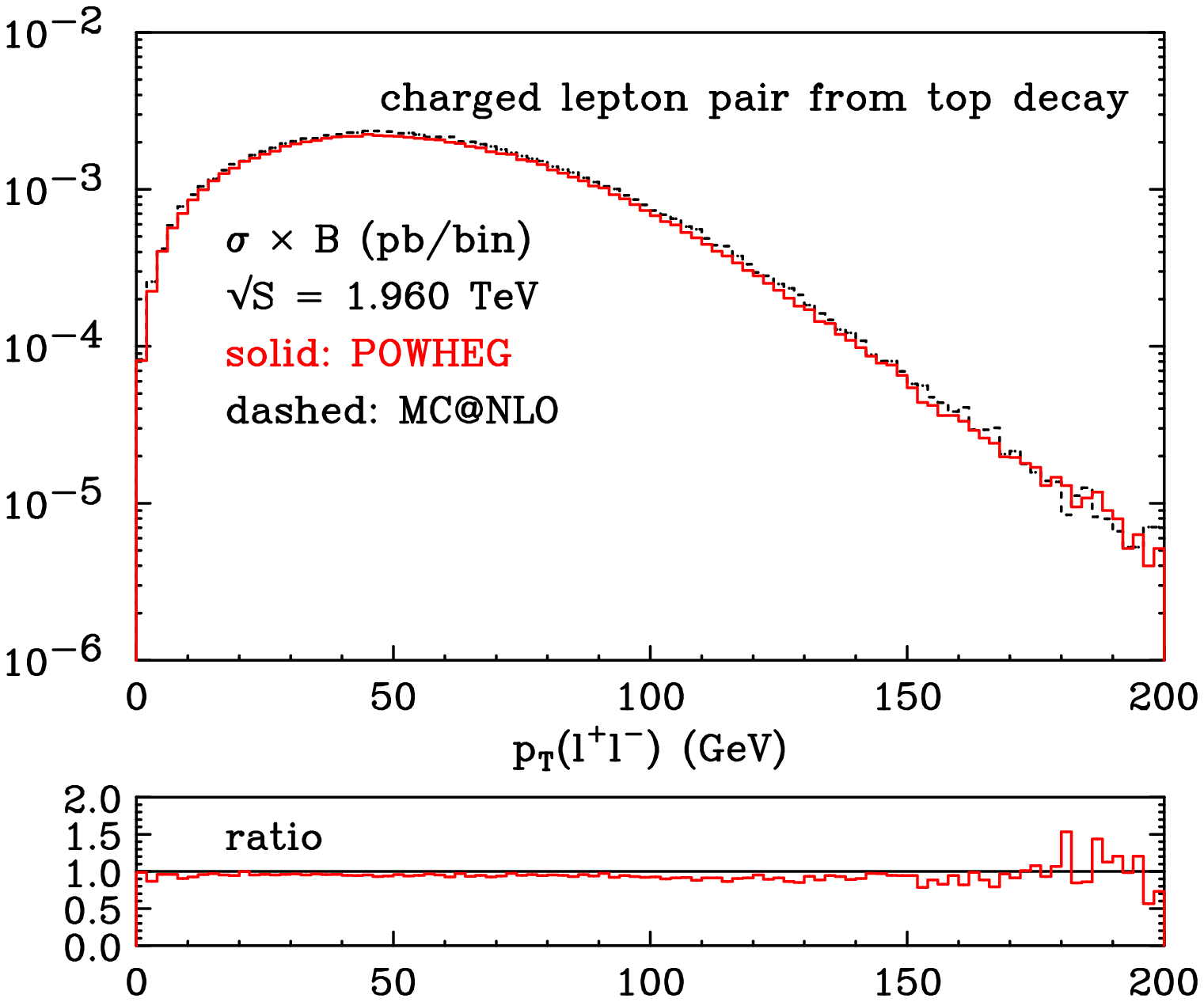,width=7cm}
\end{center}
\caption{\label{fig:toptevdec2}
Invariant mass and transverse momentum distributions
of $\ell^+\ell^-$ pairs from top decay
at the Tevatron.}
\end{figure}

\begin{figure}[ht]
\begin{center}
\epsfig{file=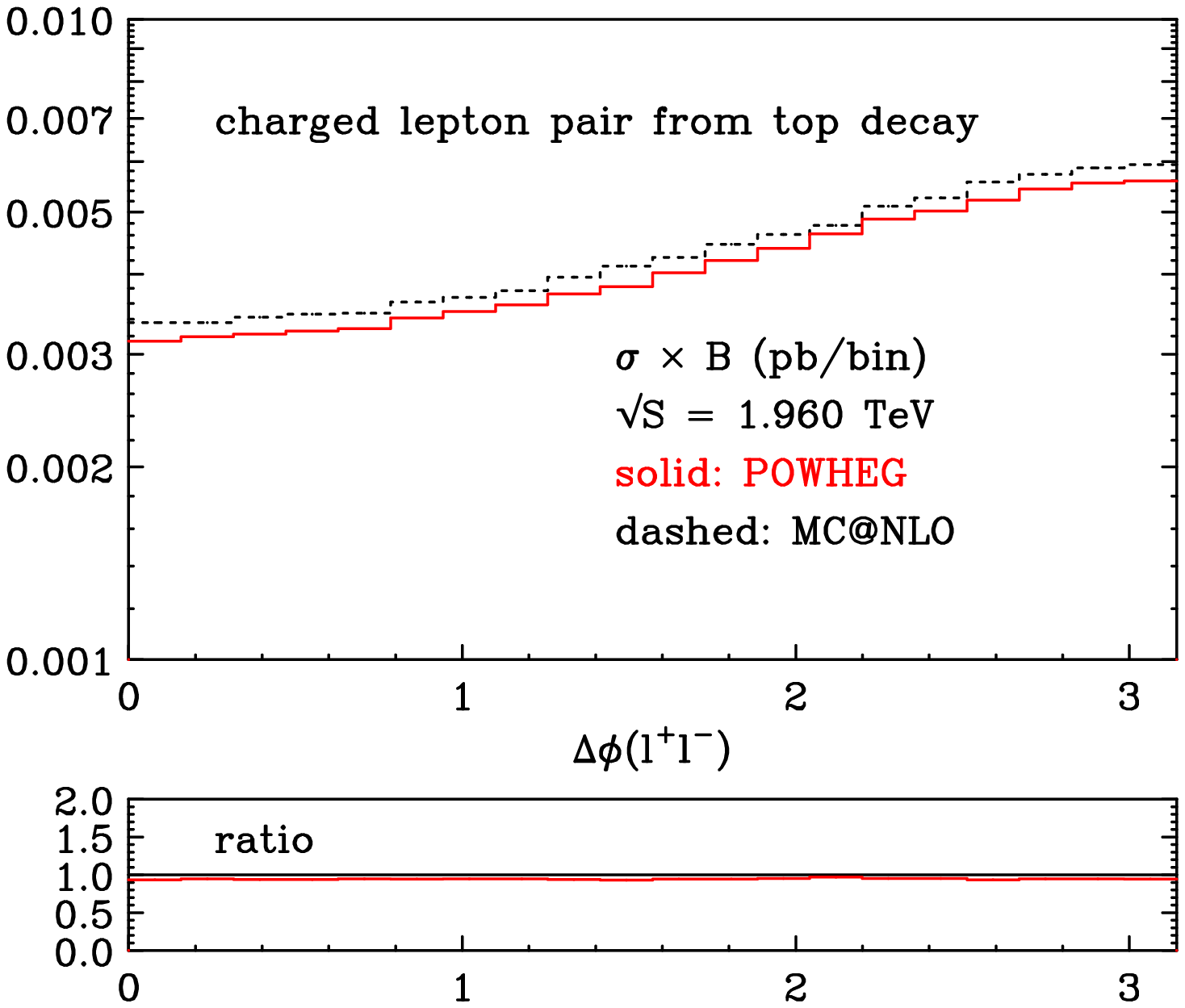,width=9cm}
\end{center}
\caption{\label{fig:toptevdec3}
Distribution of the azimuthal distance between charged leptons from top decay
at the Tevatron.}
\end{figure}

\begin{figure}[ht]
\begin{center}
\epsfig{file=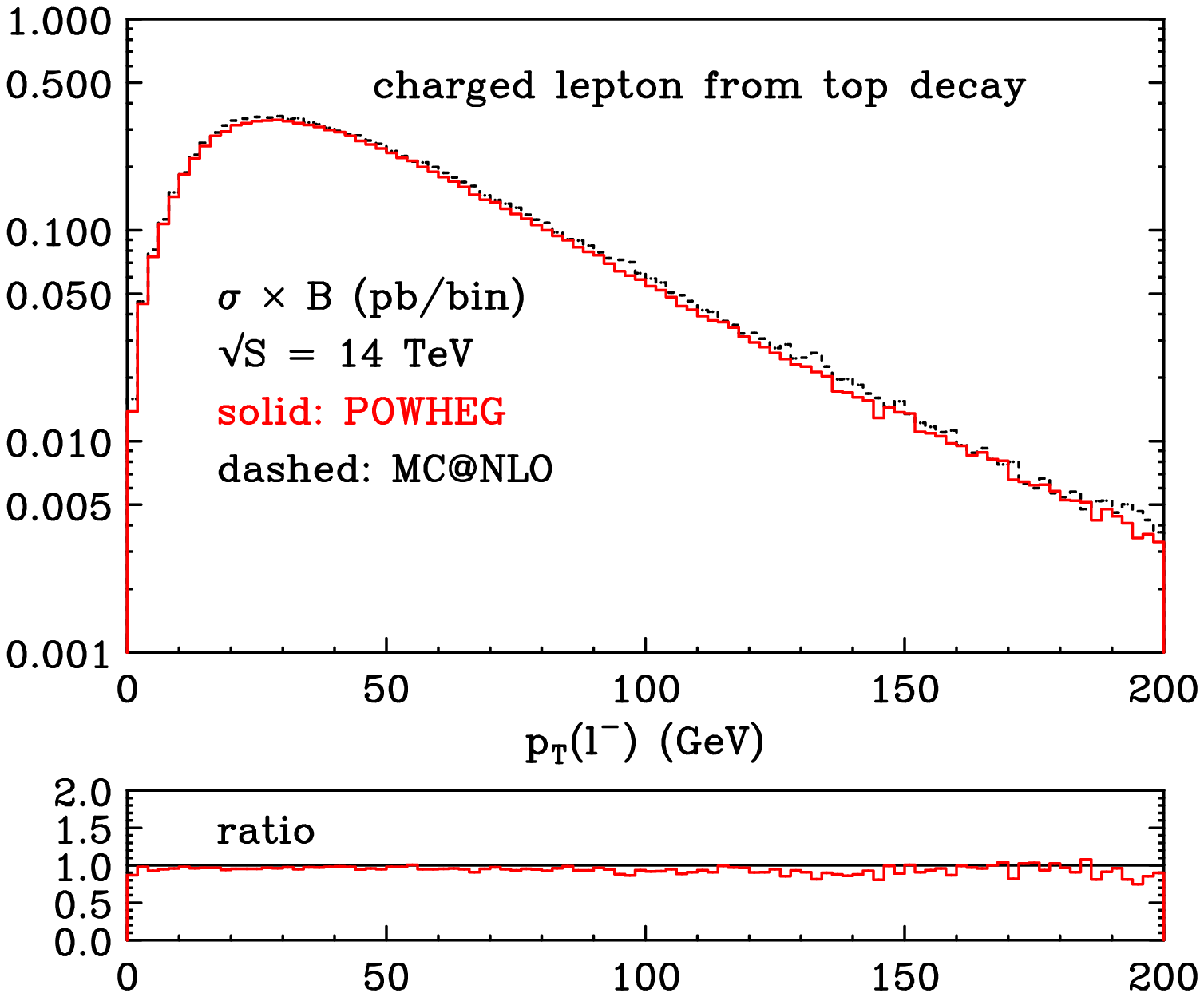,width=7cm}
\epsfig{file=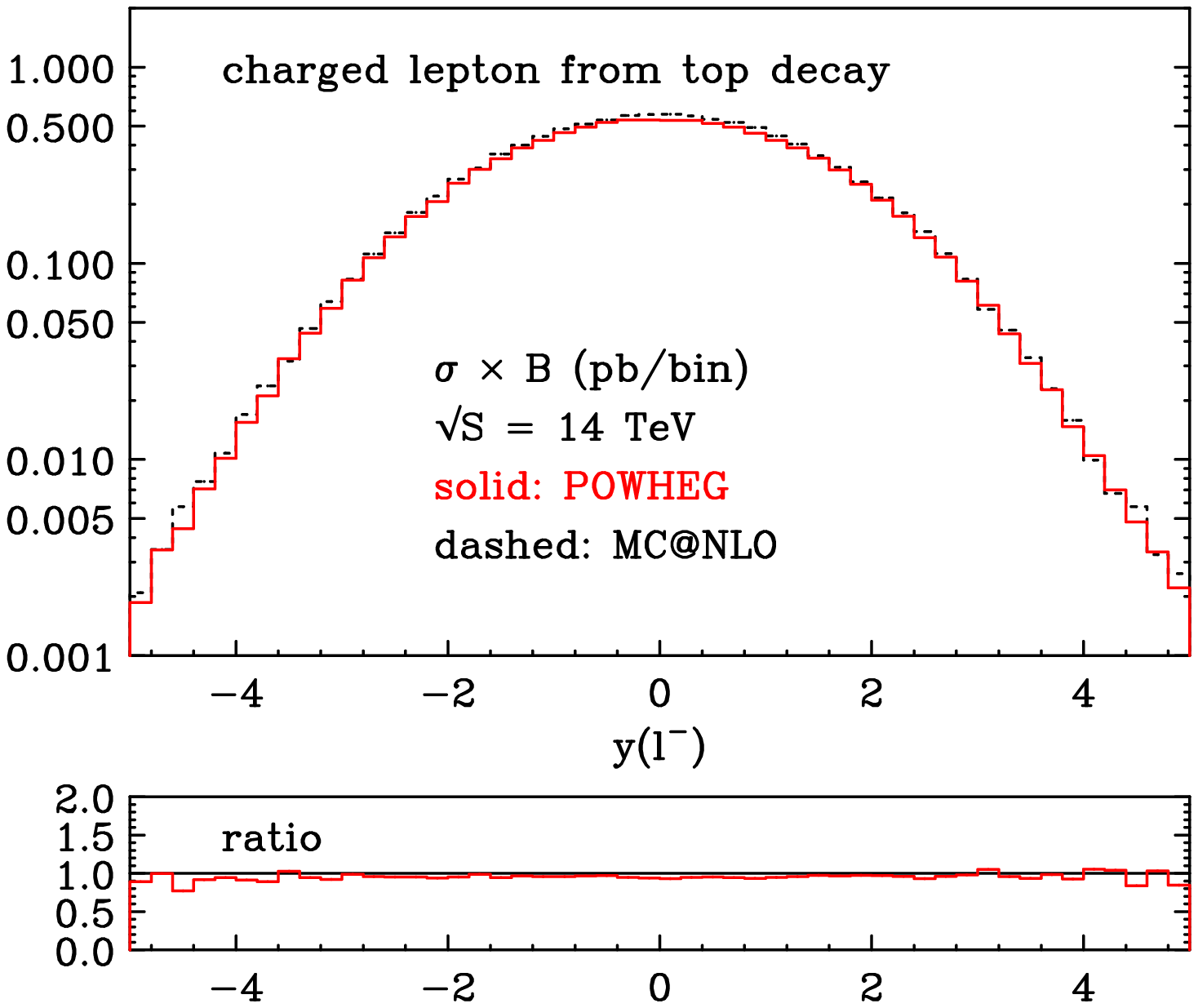,width=7cm}
\end{center}
\caption{\label{fig:toplhcdec1}
Transverse momentum and rapidity of a charged lepton from top decay
at the LHC.}
\end{figure}

\begin{figure}[ht]
\begin{center}
\epsfig{file=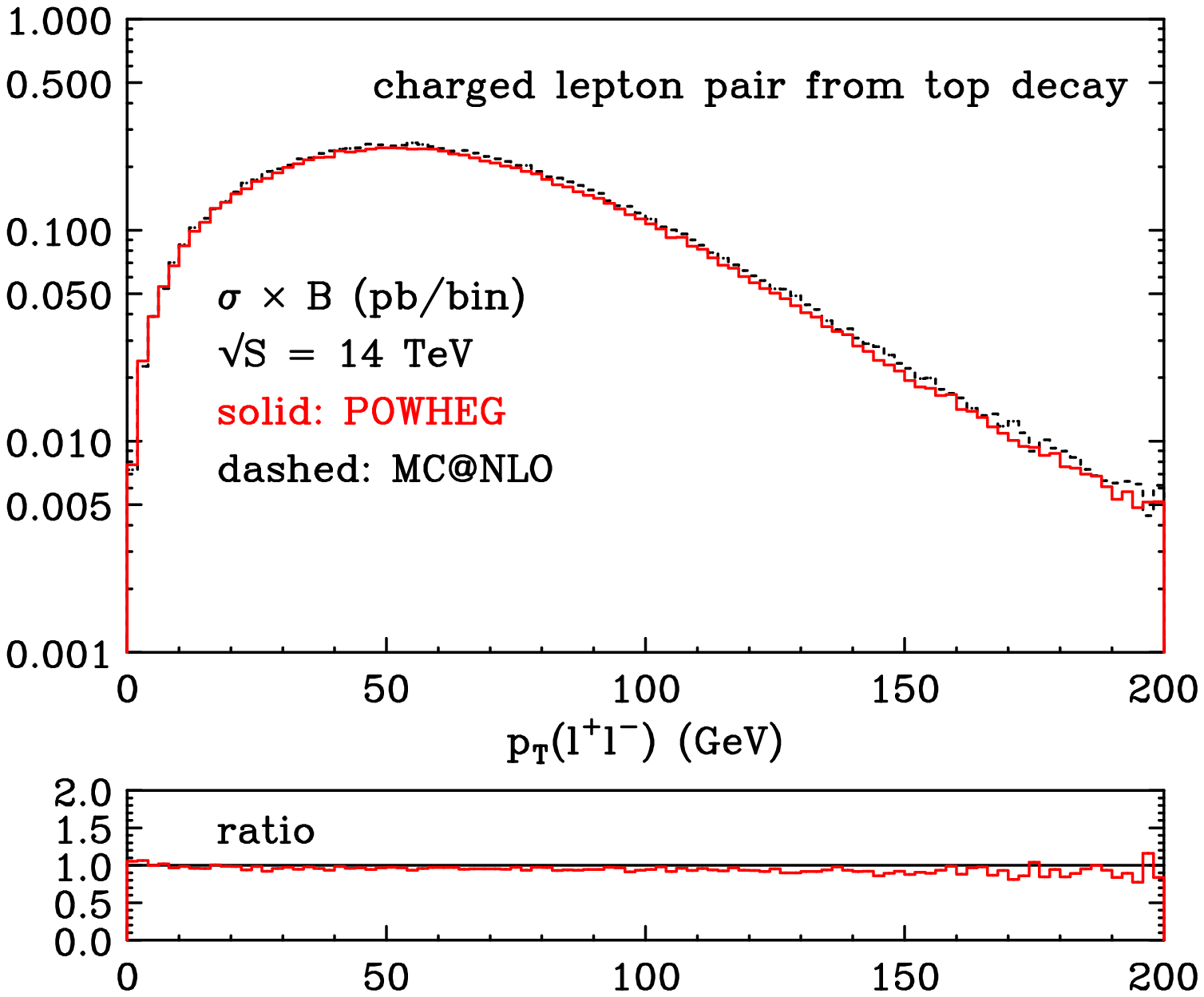,width=7cm}
\epsfig{file=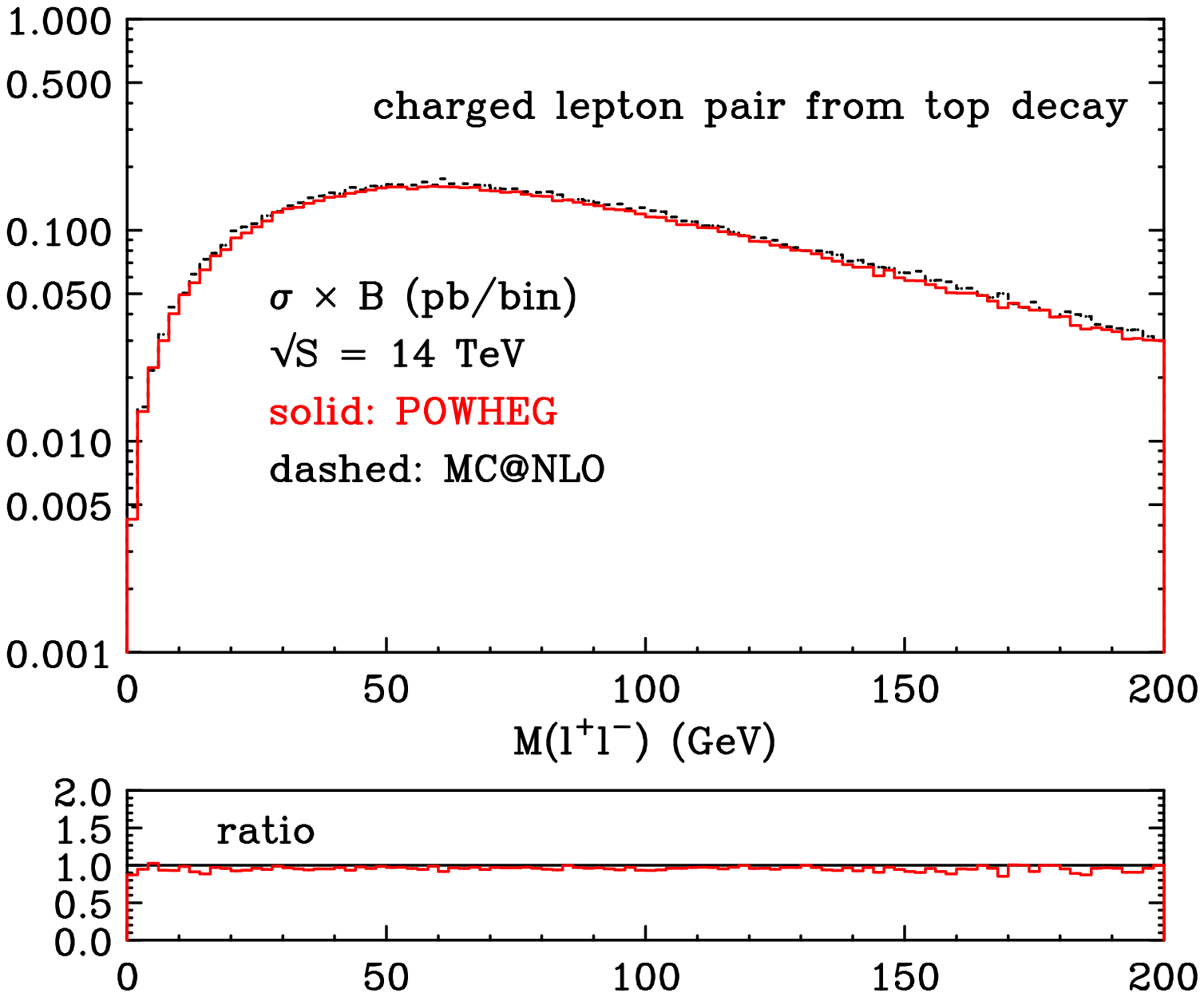,width=7cm}
\end{center}
\caption{\label{fig:toplhcdec2}
Invariant mass and transverse momentum distributions
of $\ell^+\ell^-$ pairs from top decay
at the LHC.}
\end{figure}

\begin{figure}[ht]
\begin{center}
\epsfig{file=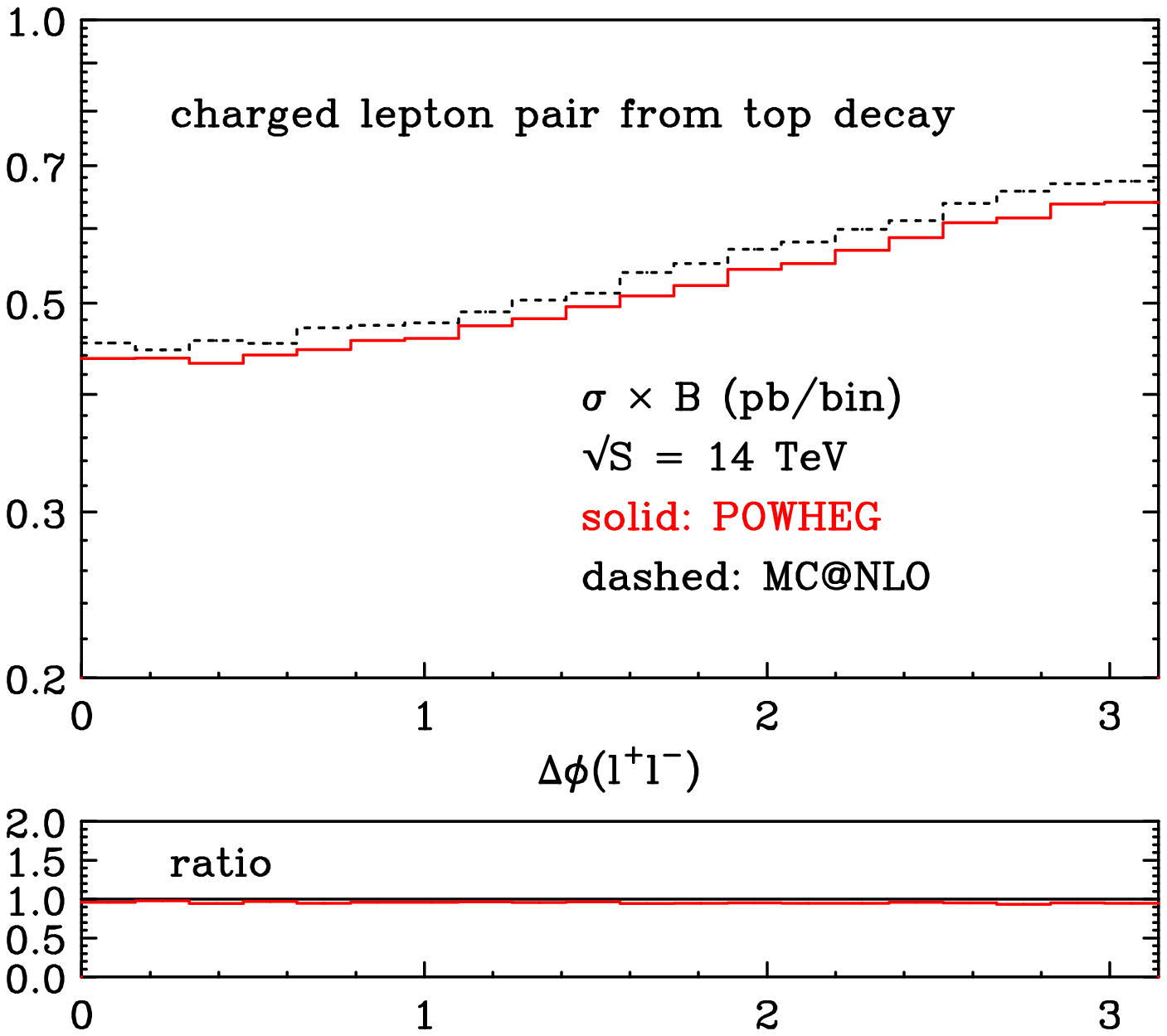,width=9cm}
\end{center}
\caption{\label{fig:toplhcdec3}
Distribution of the azimuthal distance between charged leptons from top decay
at the LHC.}
\end{figure}

We now turn to distributions of the decay products of
unstable top quarks. We consider here only the dilepton channel,
which considerably simplifies the analysis.
In fig.~\ref{fig:toptevdec1} we show two representative
single-inclusive distributions, namely
the transverse momentum and the rapidity 
of negatively-charged lepton resulting from the $\bar{t}$ decay, 
in the Tevatron configuration.
Again, both the POWHEG and the \MCatNLO\ predictions are shown, together with 
the ratio between the two results. We see that no significant difference is 
present. Similar conclusions hold for the
invariant mass and for the transverse momentum distributions
of the charged-lepton pair, fig.~\ref{fig:toptevdec2},
and for the azimuthal distance between the two charged leptons,
fig.~\ref{fig:toptevdec3}.
A similarly good agreement is obtained for the same set of observables
computed in the LHC configuration, figs.~\ref{fig:toplhcdec1}, 
\ref{fig:toplhcdec2}, and~\ref{fig:toplhcdec3}.
The only visible difference is in the overall normalization,
which is manifest in figs.~\ref{fig:toptevdec3} and~\ref{fig:toplhcdec3}.
This is due to the different choice of scales in the two computations.

We now turn to the case of bottom production. As is well known, perturbative
NLO corrections to bottom production are very large, which implies that
yet higher-order contributions are due to play a non-negligible
role. As mentioned above, we therefore expect that POWHEG 
and \MCatNLO\ will show larger discrepancies than in the case
of top production purely on the basis of fixed-order expansion. There
are, however, other sources of differences between the two approaches.
Although both codes have been interfaced to HERWIG in order to obtain
the results shown here, the logarithmically-enhanced terms beyond the 
leading one are not the same in the two approaches. 
Furthermore, if POWHEG is interfaced to
an MC based on angular-ordered evolution (such as HERWIG), standard showers 
need be supplemented by truncated showers, whose effect is that of restoring
colour coherence, which is lost because of the requirement that
the hardest radiation be always the first. Since truncated showers
are inherently soft, there are reasons to believe that their effects
are not too large. 
At present, the only study of the impact of truncated showers has
been performed in ref.~\cite{LatundeDada:2006gx}. There, a POWHEG
implementation of $e^+e^-$ annihilations into hadrons, interfaced to the
HERWIG++ Monte Carlo~\cite{Gieseke:2003hm}, was presented.
The effect of the truncated shower
was found to be small. No studies have been
performed in the case of hadron collisions.

\begin{figure}[ht]
\begin{center}
\epsfig{file=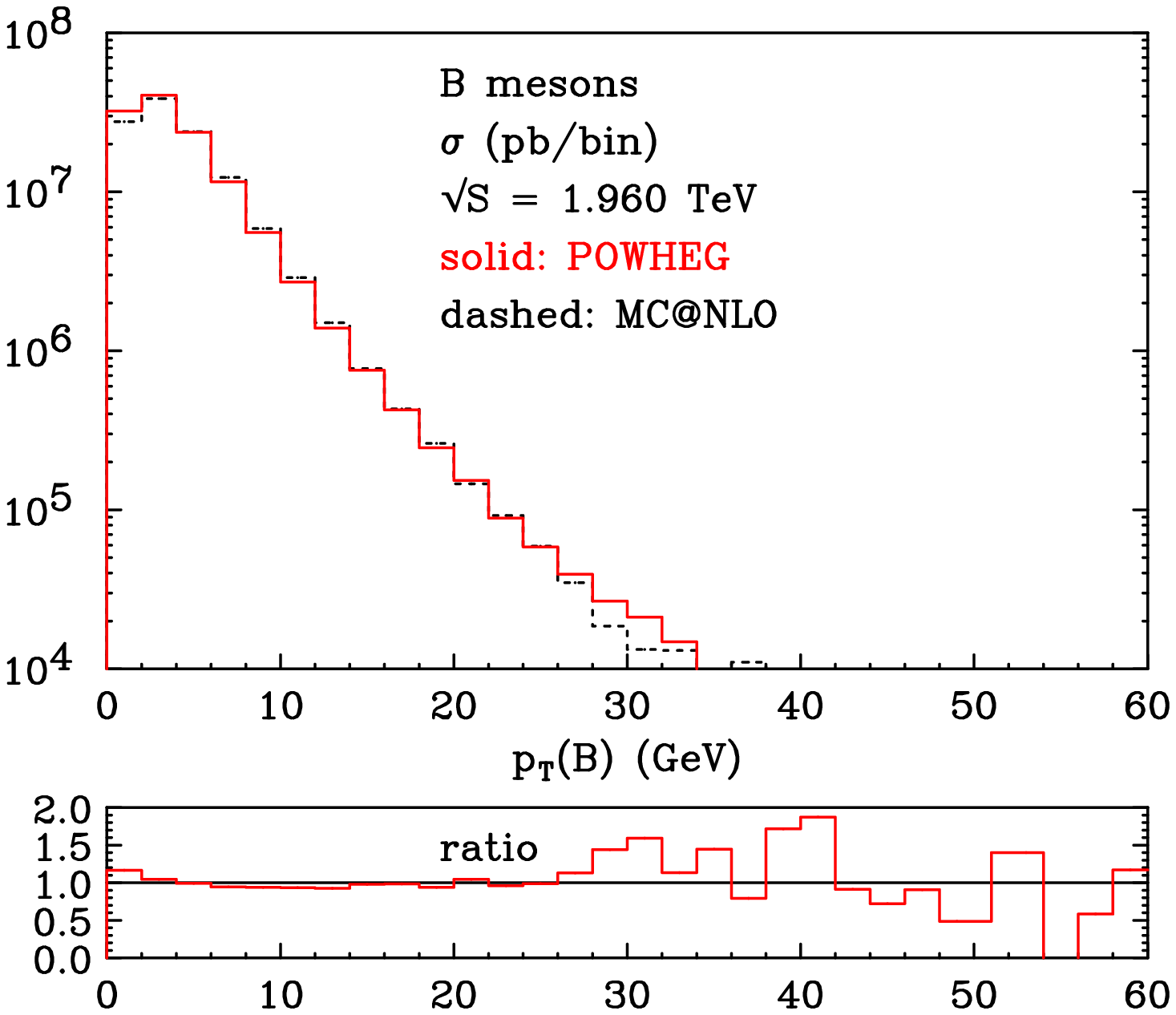,width=7cm}
\epsfig{file=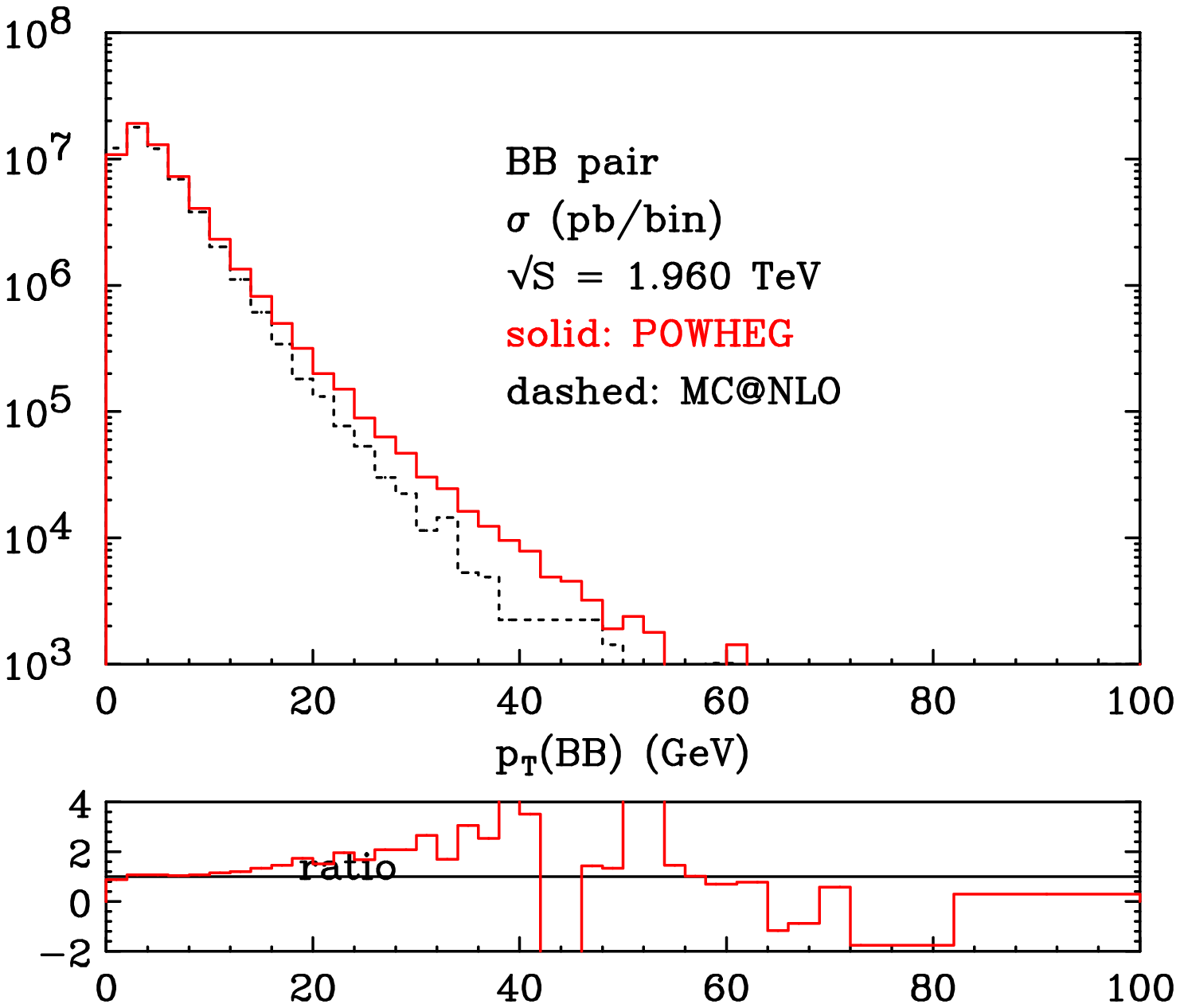,width=7cm}
\epsfig{file=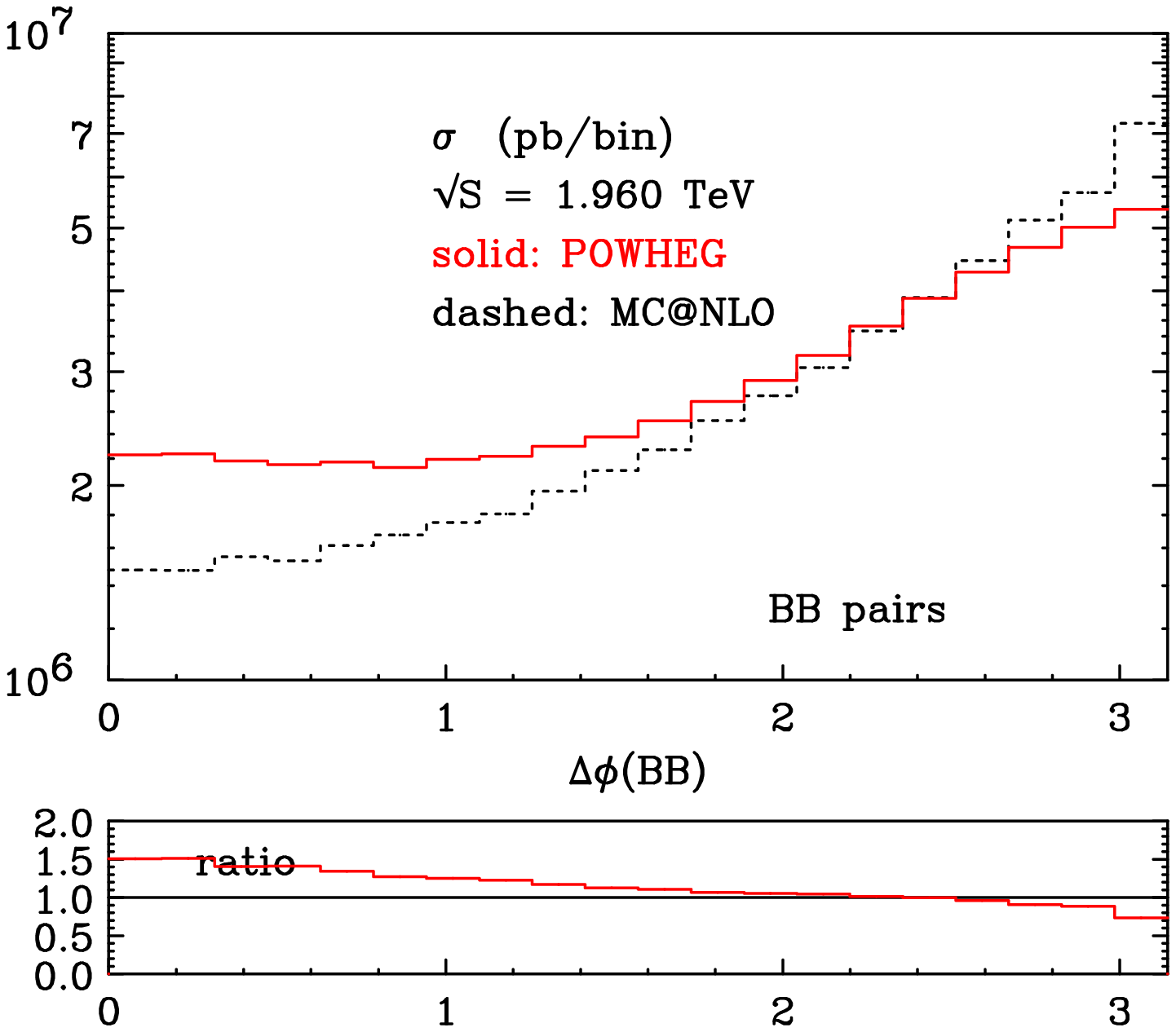,width=7cm}
\epsfig{file=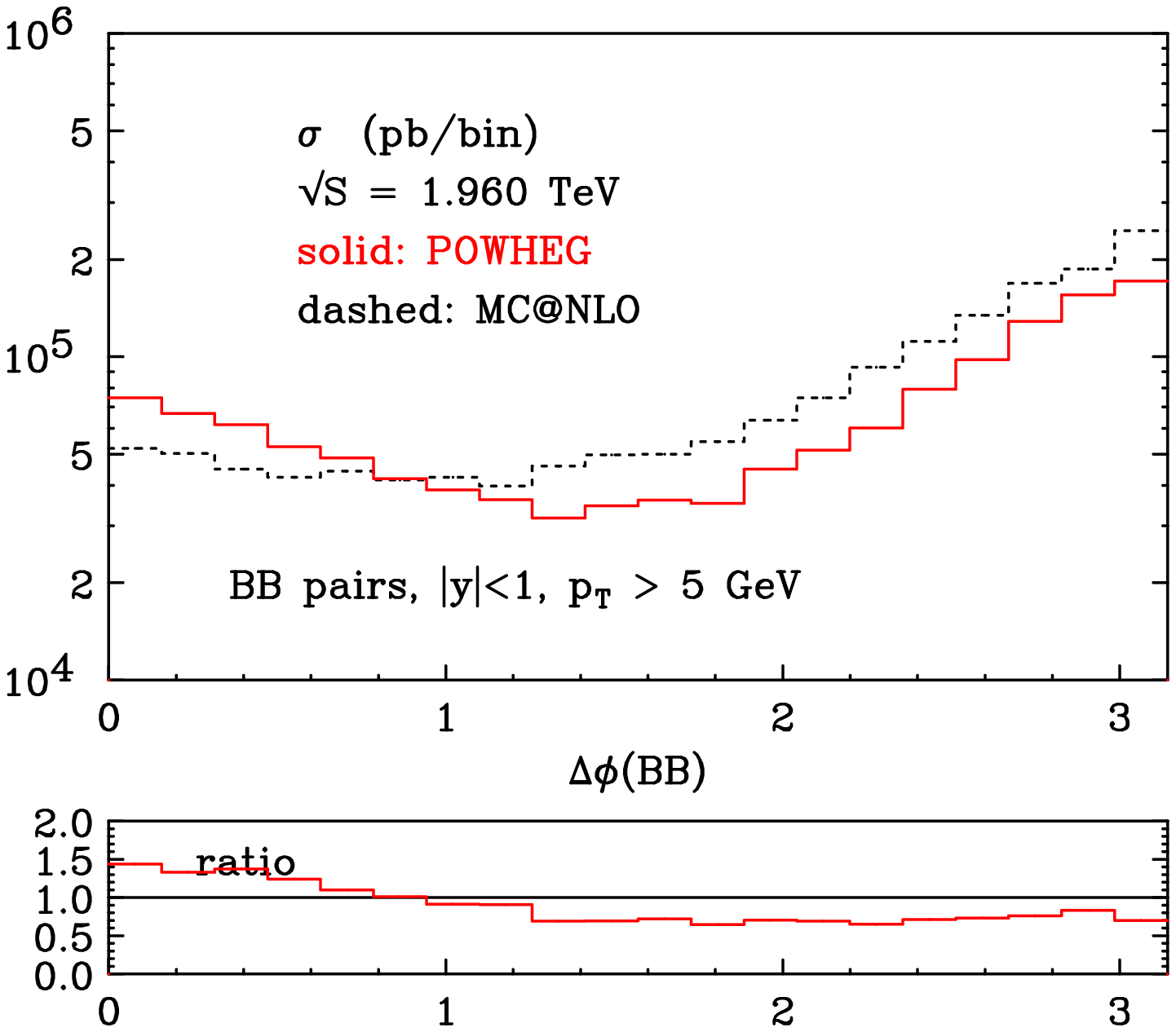,width=7cm}
\end{center}
\caption{\label{fig:bottev}
Bottom hadron distributions at the Tevatron.}
\end{figure}

In fig.~\ref{fig:bottev} we present sample comparisons between POWHEG
and \MCatNLO\ results for bottom production at the Tevatron. All observables
shown are relevant to lowest-lying $b$-flavoured meson states. We show
the single-inclusive $\pt$ (upper left pane), the pair $\pt$ (upper 
right pane), and the azimuthal distance, without (lower left pane) and
with (lower right pane) kinematic cuts; in the latter case, the cuts
$\abs{y}<1$ and $\pt>5$~GeV are applied to both the $B$'s of the pair.
The two $\pt$ distributions show a fair agreement, with POWHEG marginally
(for single inclusive $\pt$) or markedly (for the tail of the $\pt$ of the 
pair) harder than \MCatNLO. There are very clear differences in shape 
between the two azimuthal distributions. The discrepancy
tends to be smaller when cuts are applied. 
As for the $\pt$ of the pair, POWHEG gives 
harder results than \MCatNLO, which we attribute mainly to the different
treatment of hard radiation in the two formalisms.
Finally, we point out that the POWHEG code is capable of producing 
bottom and charm distributions at the LHC energy, essentially without
negatively-weighted events. 
In ref.~\cite{Frixione:2007nu} example input files are 
provided for bottom and charm production at the Tevatron and at the 
LHC, interfaced to both HERWIG and PYTHIA.

\section{Conclusions}

In this paper we have presented the implementation of heavy quark
pair production according to the POWHEG formalism, which allows an
NLO QCD computation to be matched with Parton Shower simulations.
The fortran code we have constructed can be used to predict any 
infrared safe observable in $t\bar{t}$, $b\bar{b}$, and $c\bar{c}$
production at hadron colliders.
We have compared our results with \MCatNLO\ for $t\bar{t}$ production
at the LHC and at the Tevatron, and for $b\bar{b}$ production at
the Tevatron. In the case of top production, we observe a very good 
agreement between POWHEG and \MCatNLO, for all the observables we
have considered. On the other hand, the two approaches differ
significantly for some observables in $b\bar{b}$ production,
which implies that for such low-mass quarks perturbative corrections
of order higher than next to leading are likely to play a
non-negligible role.

In general, the agreement of the \MCatNLO{} and \POWHEG{} approaches
is quite remarkable, in view of the fact that the two methods
differ considerably in several aspects, summarized below:
\begin{enumerate}
\item\label{enum:diff1}%
       The Sudakov form factors in \MCatNLO{} and \POWHEG{} are different:
      \MCatNLO{} uses HERWIG's Sudakov form factor, \POWHEG{} has its own
      (see eqs.~(\ref{eq:Deltaq}) and (\ref{eq:Deltag})).
\item\label{enum:diff2}%
      The hardest emission in \POWHEG{} carries a strong coupling
      evaluated at the $\pt$ of the emission, and a Sudakov form
      factor. On the other hand, in \MCatNLO{} only ${\mathbb S}$
      events,\footnote{See
      refs.~\cite{Frixione:2002ik,Frixione:2003ei}
      for the definition of $\mathbb S$ and $\mathbb H$ events.}
      that have all emissions entirely performed by the shower,
      have
      these features. The $\mathbb{H}$ events are evaluated at the scale
      of the hard process, and they carry no Sudakov damping for small
      transverse momenta. This difference may show up for relatively low
      transverse momentum. We remind the reader that $\mathbb{H}$ events
      can have negative weight, so that it is difficult to understand
      in which direction this difference affects the results.
\item\label{enum:diff3}%
      The \POWHEG{} approach lacks the truncated showers.
\item\label{enum:diff4}%
      Subleading terms in the shower may differ in the two approaches, due
      to the reshuffling of the splitting processes in the shower illustrated in
      ref.~\cite{Nason:2004rx}.
\item\label{enum:diff5}%
      The $\mathbb{H}$ events in \MCatNLO{} may be followed by radiation,
      generated by the shower, with a $\pt$ harder than the $\pt$
      of the $\mathbb{H}$ event.
      In \POWHEG{} harder emissions from the shower are always
      vetoed.
\end{enumerate}
Because of the many differences,
it is also difficult at this stage to understand what causes the
differences in the distributions we have presented. Here we just
make a few speculations about the possible origin of the differences,
that should only be taken as hints for further studies.
First, we look at top production.
We see there that the inclusive $\pt$
spectrum,
the mass of the pair and the $\pt$ of the pair
differ in the very small $\pt$ or $m(tt)$ region, \POWHEG{} being generally
higher (see figs. \ref{fig:toptevsta1} to \ref{fig:toplhcsta2}).
The lack of soft-truncated showers in \POWHEG{}
(item \ref{enum:diff3} of the above list)
could possibly cause this
effect. On the other hand, soft radiation is also treated differently as
far as the hardest emission is concerned, as specified in item~\ref{enum:diff2}
of the above list.
The fact that the difference goes in the opposite way (i.e. \POWHEG{}
is below \MCatNLO{}) for the transverse momentum of the top pair at LHC
also shows that the lack of truncated showers (that would lower the \POWHEG{}
distribution) cannot be the whole answer.
A second effect we notice is the considerable
difference in the azimuthal distance
of the bottom pair (see fig. \ref{fig:bottev})
especially in the region where the two heavy mesons
are near in azimuth. Here, the lack of soft-truncated showers in \POWHEG{}, or 
differences in subleading
shower effects in the two methods, could yield a different degree of
smearing of the azimuthal distance. The \POWHEG{} result seems to have
more smearing than the \MCatNLO{} result. The lack of soft-truncated showers
is likely to have the opposite effect. Also, the faster rise
for small azimuthal difference is unlikely to be due to shower effects
in general, and would suggest to look for effects in the hard radiation
mechanism. Thus, items \ref{enum:diff2}, \ref{enum:diff4} and \ref{enum:diff5}
may be responsible for these differences.
The third effect we consider is the harder tail of the $\pt$ distribution
of bottom pairs in \POWHEG{}. This can only be ascribed to genuine
higher order effects in the hard emission, as may arise from
items \ref{enum:diff1} and \ref{enum:diff2}.\footnote{In a yet unpublished
revision of the \MCatNLO{} code the $\pt$ spectra for $B$ production
turn out to be harder than the ones shown here.
The azimuthal distance distributions
do not differ from those presented here.}

\section*{Acknowledgement}
We thank Rachid Guernane for testing our code
and suggesting speed improvements.

\providecommand{\href}[2]{#2}\begingroup\raggedright\endgroup

\end{document}